\documentclass[12pt]{iopart}

\usepackage{iopams}
\usepackage{setstack}
\usepackage{subfig}
\usepackage{graphicx}
\usepackage{float}
\usepackage{cite}
\usepackage[labelfont=bf, textfont=it]{caption}
\usepackage{hyperref}
\usepackage{xcolor}
\usepackage{tabularx}
\usepackage{colortbl}
\usepackage{multirow}
\usepackage{bm}
\usepackage{booktabs}
\usepackage{amssymb}
\usepackage{makecell}

\definecolor{Gray}{gray}{0.85}
\definecolor{LightCyan}{rgb}{0.88,1,1}

\newcolumntype{a}{>{\columncolor{Gray}}c}

\newcommand{\secref}[1]{\S\ref{#1}}

\begin{document}

\title{On electromagnetic turbulence and transport in STEP}

\author{M. Giacomin$^{1,2}$, D. Kennedy$^3$, F. J. Casson$^3$, Ajay C. J.$^4$, D. Dickinson$^1$, B. S. Patel$^3$ and C. M. Roach$^3$}

\address{$^1$York Plasma Institute, University of York, York, YO10 5DD, United Kingdom}
\address{$^2$Dipartimento di Fisica ``G. Galilei'', Università degli Studi di Padova, Padova, Italy}
\address{$^3$Culham Centre for Fusion Energy, Abingdon, OX14 3DB, United Kingdom}
\address{$^4$Centre for Fusion, Space and Astrophysics, Department of Physics, University
of Warwick, Coventry, CV4 7AL, United Kingdom}

\ead{maurizio.giacomin@york.ac.uk}

\begin{abstract}

In this work, we present first-of-their-kind nonlinear local gyrokinetic simulations of electromagnetic turbulence at mid-radius in the burning plasma phase of the conceptual high-$\beta$, reactor-scale, tight-aspect-ratio tokamak STEP (Spherical Tokamak for Energy Production). A prior linear analysis in D. Kennedy \textit{et al.} 2023 Nucl. Fusion 63 126061 reveals the presence of unstable hybrid kinetic ballooning modes, where inclusion of the compressional magnetic field fluctuation, $\delta B_{\parallel}$, is crucial, and subdominant microtearing modes are found at binormal scales approaching the ion-Larmor radius.
Local nonlinear gyrokinetic simulations on the  selected surface in the central core region suggest that hybrid kinetic ballooning modes can drive large turbulent transport, and that there is negligible turbulent transport from subdominant microtearing modes when hybrid kinetic ballooning modes are artificially suppressed (through the omission of $\delta B_{\parallel}$). 
Nonlinear simulations that include perpendicular equilibrium flow shear can saturate at lower fluxes that are more consistent with the available sources in STEP.
This analysis suggests that hybrid kinetic ballooning modes could play an important role in setting the turbulent transport in STEP, and possible mechanisms to mitigate turbulent transport are discussed. \textcolor{black}{Increasing the safety factor or the pressure gradient strongly reduces turbulent transport from hybrid kinetic ballooning modes in the cases considered here.}  Challenges of simulating electromagnetic turbulence in this high-$\beta$ regime are highlighted. In particular the observation of radially extended turbulent structures in the absence of equilibrium flow shear motivates future advanced global gyrokinetic simulations that include $\delta B_\parallel$.    

\end{abstract}

\section{Introduction}
\label{sec:introduction}

Understanding and predicting turbulence in the core of next-generation spherical tokamaks (STs) is critical for the optimisation of their performance. 
In order to be economically competitive, ST power plants require a high ratio, $\beta$, of thermal pressure to magnetic pressure in the tokamak core. At high $\beta$, turbulence is expected to become more electromagnetic in nature, and to be influenced by kinetic ballooning modes (KBMs) and microtearing modes (MTMs), as frequently reported for STs (see \cite{kaye2021} and references therein). 
Whilst gyrokinetic (GK) simulations have thus far proven to be a reliable tool for modelling turbulent transport in predominantly electrostatic regimes at low $\beta$ in conventional aspect ratio tokamaks, obtaining saturated nonlinear simulations of plasmas at higher $\beta$ with unstable KBMs and MTMs has proved more challenging~\cite{patelthesis,dickinson2022}. 

The UK Spherical Tokamak for Energy Production (STEP) programme is developing a compact prototype power plant design based on a high-$\beta$ ST, with the primary goal of generating net electric power $P_\mathrm{el} > 100$ MW from fusion~\cite{wilson2020,meyer2022,patel2021,morris2022}. The first phase of this ambitious programme is to provide a conceptual design of the STEP prototype plant and a reference equilibrium for the preferred flat-top operating point. 
Various reference plasma scenarios for STEP~\cite{meyer2022,tholerus2024} have been developed using the integrated modelling suite JINTRAC~\cite{romanelli2014}, with plasma transport calculated using the Bohm-gyro-Bohm model~\cite{Erba1997} with transport coefficients rescaled (i) to give an energy confinement time enhancement factor $H_{98}$ = 1.3, over the ITER98,y2 \cite{WAKATANI_NF1999} confinement scaling law, and (ii) to give dominant electron heat transport (as frequently observed in STs \cite{kaye2021}).

We will focus here on the STEP reference scenario STEP-EC-HD \footnote{More specifically we use STEP-EC-HD=v5 with SimDB UUID: 2bb77572-d832-11ec-b2e3-679f5f37cafe, Alias: smars/jetto/step/88888/apr2922/seq-1} (where EC stands for Electron Cyclotron heating and current drive, and HD stands for High Density), which is designed to deliver a fusion power $P_{\mathrm{fus}} = 1.8$~GW. The global parameters for this plasma flat-top operating point are reported in \cite{kennedy2023a}, \textcolor{black}{while details on the JINTRAC modelling of the chosen STEP flat-top operating point are reported in \cite{meyer2022,tholerus2024}}.
At this STEP flat-top operating point, the heat source is mainly from fusion $\alpha$-particles and from electron-cyclotron heating, while the particle source is from pellet injection that provides 7.4 $\times 10^{21}$ particles/s\footnote{In modelling this flat-top operating point the particle confinement time $\tau_p \sim 4.5 \tau_E$ is assumed consistent with typical findings in JET. Assuming the same confinement time for helium ash gives a saturated helium abundance of about 9\%. Higher helium confinement would degrade fusion performance due to the core accumulation of helium ash.}.

This paper seeks to progress the important objective of testing the transport and confinement assumptions used to develop STEP scenarios against first-principles-based local GK simulations; these simulations are expected to be considerably more challenging than for conventional aspect ratio tokamaks at low $\beta$, because of the expected electromagnetic nature of the turbulence in high-$\beta$ STs. The main goals of this work are (a) to model plasma turbulence in a STEP flat-top operating point with the best currently available tools;  (b) to assess the compatibility of the predicted transport with the anticipated sources in STEP; (c) to explore strategies for optimising transport in the conceptual STEP design; and (d) to identify modelling challenges and suggest future directions for tackling them.

The nonlinear simulations presented here are carried out using the local gyrokinetic code CGYRO~\cite{candy2016} (commit \texttt{399deb4c}). 
All simulations include three species (electrons, deuterium and tritium) and neglect the impacts of impurities and fast $\alpha$ particles, which will be assessed in future work.

The paper is organised as follows. Firstly the reference local equilibrium used throughout is introduced in \secref{sec:localmustab} and its local microstability properties are summarised: this is the $q=3.5$ surface close to mid-radius in STEP-EC-HD.  Then \secref{sec:hybrid KBM instability} presents fully electromagnetic (i.e. including $\delta B_{\parallel}$) local nonlinear gyrokinetic simulations of the hybrid-KBM dominated turbulence, and simulations are performed for various assumed values of the perpendicular equilibrium flow shearing rate. In \secref{sec: sensitivity to local parameters}, we explore possible mechanisms to mitigate large turbulent fluxes that are driven by the hybrid-KBMs and we assess sensitivity of the turbulent transport to local pressure gradient, $\beta$ and safety factor, in order to inform the design of future STEP flat-top operating points.
Simulations of isolated turbulence from the subdominant MTM instability are performed in \secref{sec:subdominant MTM instability} in a situation where hybrid-KBMs are artificially suppressed by removing $\delta B_{\parallel}$.  
Finally, a summary of findings is presented in \secref{sec:conclusions}.

\section{Local equilibrium and microstability at \texorpdfstring{$\boldsymbol{q=3.5}$}{TEXT} in STEP-EC-HD}
\label{sec:localmustab}

\begin{figure}
    \centering
    \subfloat[]{\includegraphics[height=0.32\textheight]{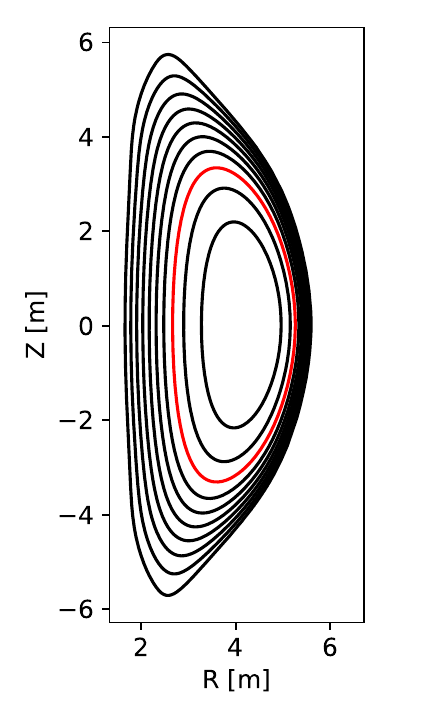}}\quad
    \subfloat[]{\includegraphics[height=0.32\textheight]{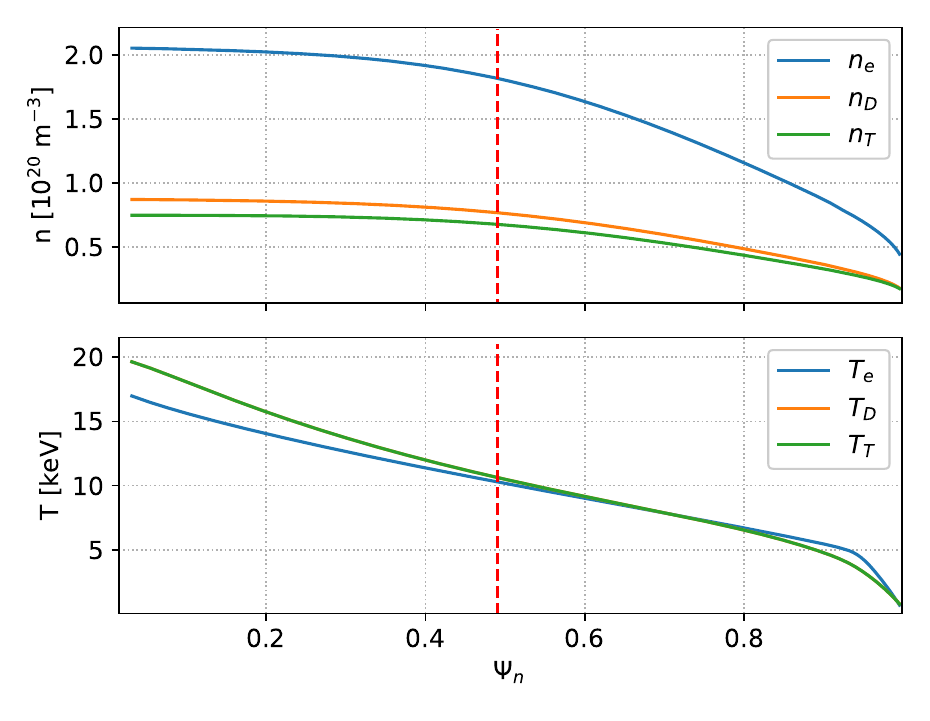}}
    \caption{Profiles from the STEP-EC-HD flat-top equilibrium: (a) contour plot of the poloidal magnetic flux, with the red contour denoting the $q=3.5$ flux surface corresponding to normalised poloidal flux $\Psi_n=0.49$; (b) density and temperature profiles of electrons, deuterium and tritium as functions of $\Psi_n$.}
    \label{fig:equilibria}
\end{figure}

The local gyrokinetic analysis in this paper is carried out at the $q=3.5$ surface of the STEP-EC-HD flat-top equilibrium.  This surface is highlighted in red in Figure~\ref{fig:equilibria}, which also shows radial profiles of the three plasma species included in the gyrokinetic simulations; electrons, deuterium and tritium\footnote{The source reference equilibrium includes impurities and helium, but our GK calculations neglect these species and scale the deuterium and tritium density by a factor close to unity to restore quasi-neutrality.}. 
A Miller parameterisation \cite{miller1998} was used to model the local magnetic equilibrium geometry, and the shaping parameters were fitted to the chosen surface using \texttt{pyrokinetics} \cite{pyrokinetics}, which is a python library developed to facilitate pre- and post-processing of gyrokinetic analysis. Table~\ref{tab:equilibrium} provides the values of various local equilibrium quantities on this surface, including: magnetic shear, $\hat{s}$; normalised minor radius, $\rho/a$; elongation and its radial derivative, $\kappa$ and $\kappa'$; triangularity and its radial derivative, $\delta$ and $\delta'$; radial derivative of the Shafranov shift, $\Delta'$; temperature $T$ and density $n$ for electrons, deuterium and tritium, and their normalised inverse gradient scale lengths.

\begin{table}
\centering
\begin{tabular}{|a|c|a|c|}
\hline
\textbf{Parameter} & \textbf{Value} & \textbf{Parameter} & \textbf{Value}\\
\hline
$\Psi_n$ & 0.49 & ${n_e\,[10^{20}\mathrm{m}^{-3}]}$ & 1.81  \\
\hline
$q$ & 3.5 & ${T_e\, [\mathrm{keV}]}$ & 10.3\\
\hline
$\hat{s}$ & 1.2 & $\rho_s$ [mm] & 5.2\\
\hline
$\kappa$ & 2.56 & \textcolor{black}{$(a/c_s)\nu_{ee}$} & 0.03 \\
\hline
$\kappa'$ & 0.06 &  $n_D/n_e$ & 0.53\\
\hline
$\delta$ & 0.29 &  $n_T/n_e$ & 0.47\\
\hline
$\delta'$ & 0.46 &  $T_D/T_e$ & 1.03\\
\hline
$\Delta'$ & -0.40 &  $T_T/T_e$ & 1.03\\
\hline
$\beta_e$ & 0.09 &  $a/L_{n_e}$ & 1.03\\
\hline
$\beta'$ & -0.48 &  $a/L_{n_D}$ &  1.06 \\
\hline
$r$ [m] & 1.3 & $a/L_{n_T}$ & 0.99 \\
\hline
$R$ [m] & 4.0 & $a/L_{T_e}$ &  1.58 \\
\hline
$A_\mathrm{surf}$\,[m$^2$] & 370 & $a/L_{T_D}$ & 1.82 \\
\hline
$P_{\mathrm{surf}}$\,[MW] & 500 & $a/L_{T_T}$& 1.82\\
\hline
${B_0\, [\mathrm{T}]}$ & 2.8 & &\\
\hline
\end{tabular}
\caption{Local parameters at $q=3.5$ in the STEP-EC-HD flat-top. Parameters not defined in the text include:  $\beta_e=2\mu_0n_eT_e/B_0^2$; minor radius, $a[\mathrm{m}] = 2$, is the half-diameter at the equatorial midplane of the last closed flux-surface; local minor radius, $r$; \textcolor{black}{$\nu_{ee}$ is the electron collision frequency}; $X^{\prime}=\frac{dX}{d\rho}$ denotes the derivative of quantity $X$ with respect to the flux label $\rho=r/a$; major radius $R$; surface area $A_\mathrm{surf}$; total heating power crossing the surface $P_\mathrm{surf}$; density $n_s$ and temperature $T_s$ for species $s$; and the normalised inverse gradient scale length for quantity $X$, $a/L_X = \frac{a}{X} \frac{\mathrm{d}X}{\mathrm{d}\rho}$.}
\label{tab:equilibrium}
\end{table}

\begin{figure}
    \centering
    \subfloat[]{\includegraphics[width=0.48\textwidth]{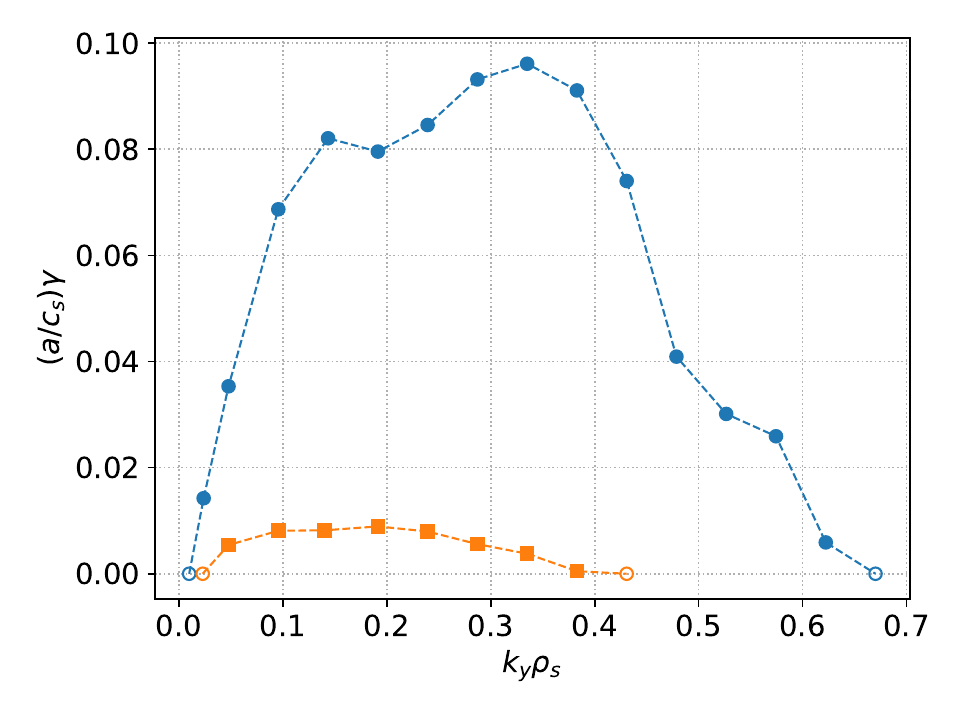}}\,
    \subfloat[]{\includegraphics[width=0.48\textwidth]{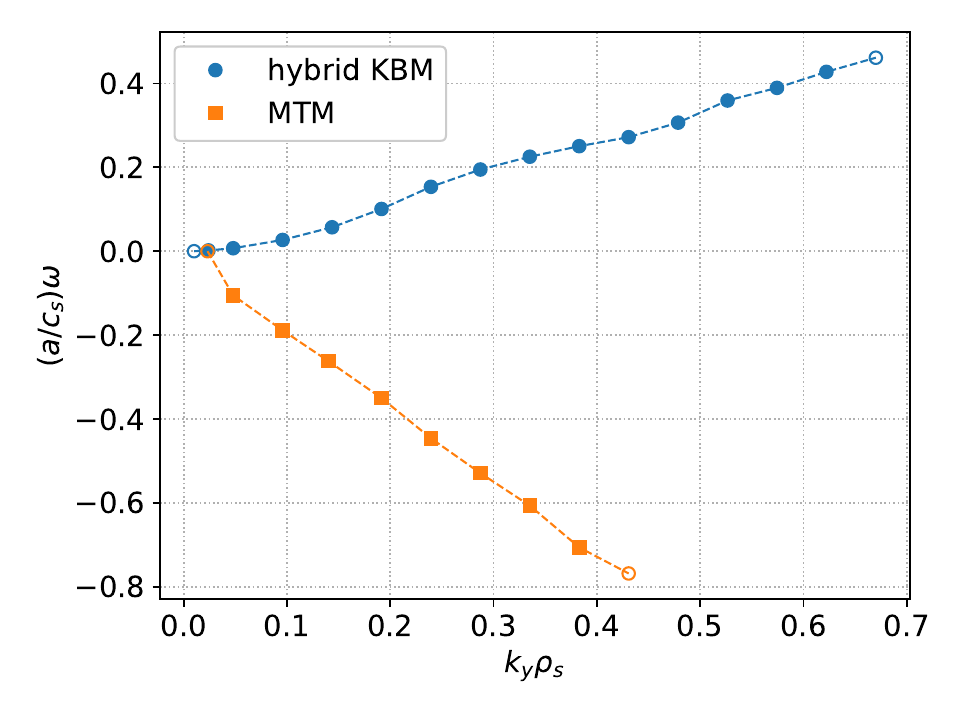}}
    \caption{Growth rate (a) and mode frequency (b) as a function of $k_y\rho_s$ from linear simulations of the dominant (hybrid-KBM) and subdominant (MTM) instabilities of STEP-EC-HD at the radial surface corresponding to $q=3.5$ (see \cite{kennedy2023a} for details on the linear analysis). Stable modes are shown as open markers.}
    \label{fig:linear}
\end{figure}

Local linear GK analysis on this $q=3.5$ surface of STEP-EC-HD is reported in \cite{kennedy2023a}, and reveals the presence, at ion Larmor radius scales, of a dominant  hybrid-KBM instability that has significant linear drive contributions from trapped electrons and electromagnetic terms. A subdominant MTM is also unstable at ion binormal scales, and no unstable mode is observed at electron Larmor radius scales. Figure~\ref{fig:linear}  shows the linear growth rate, $\gamma$, and mode frequency, $\omega$, of the dominant and sub-dominant modes as functions of the normalised binormal wave-number $k_y\rho_s$, where $\rho_s$ denotes the deuterium Larmor radius (evaluated using the electron temperature) $\rho_s = c_s/\Omega_D$, $c_s=\sqrt{T_e/m_D}$ is the deuterium sound speed, $\Omega_D=eB_0/m_D$ is the deuterium cyclotron frequency, \textcolor{black}{$m_D$ is the deuterium mass}, and $B_0$ is the toroidal magnetic field at the centre of the flux surface. Full details on the linear physics are reported in \cite{kennedy2023a}, \textcolor{black}{where the linear properties of the hybrid-KBM are characterised in detail.} The work in \cite{kennedy2023a} points out the critical role played by compressional magnetic fluctuations, $\delta B_\parallel$.  If $\delta B_{\parallel}$ is artificially excluded from calculations the hybrid-KBM is entirely stabilised, while the linear properties of the MTM are unaffected. Here we exploit this in order to model the turbulent transport that might be expected to be driven by these two classes of mode if they could be modelled independently.

Whilst in this paper the analysis is restricted to $q=3.5$, we expect general results and trends to be relevant across a range of surfaces in STEP-EC-HD, which have been shown in \cite{kennedy2023a} to share qualitatively similar linear microstability properties. Further details on the local equilibrium are reported in \cite{kennedy2023a}.

\section{Nonlinear simulations including \texorpdfstring{$\boldsymbol{\delta B_{\parallel}}$: hybrid-KBM turbulence}{TEXT}}
\label{sec:hybrid KBM instability}

The focus here is to analyse turbulence driven by  hybrid-KBMs.  This requires fully electromagnetic nonlinear gyrokinetic simulations that include perturbations in electrostatic potential $\delta \phi$, parallel magnetic vector potential $\delta A_{\parallel},$ and parallel magnetic field $\delta B_{\parallel}$.  

\subsection{Numerical resolution}
\label{sec:resolution}

The CGYRO simulations have generally been performed using grid sizes\footnote{CGYRO discretises velocity space using Laguerre and Legendre spectral methods in $v$ and $\xi$, respectively (see \cite{candy2016} for details), where $\xi = v_{\parallel}/v$  with $v_{\parallel}$ the velocity component parallel to the equilibrium magnetic field direction.} in the parallel, binormal, radial, energy and pitch-angle dimensions, $(n_\theta, n_{k_y}, n_{k_x}, n_v, n_\xi) = (32,96,128,8,32)$, and using the Sugama collision model~\cite{sugama2009}. The binormal $k_y$ grid spans $0 \leq k_y \rho_s \leq 0.95$, so includes the full range of linear instability and linearly stable grid-points at both high and low $k_y$ (see Figure~\ref{fig:linear}).
The radial box size $L_x = j/(\hat{s} k_{y,\mathrm{min}})$ is set to accommodate exactly eight rational surfaces of the lowest finite toroidal mode number by choosing the integer $j=8$, motivated by the numerical convergence study\footnote{Small values of $\Delta k_x$ were also required in local GK simulations of ITG turbulence in MAST in the presence of equilibrium flow shear.  While the radial domain of the flux-tube was larger than MAST minor radius, the simulated turbulence properties were nevertheless found to be in remarkably close agreement with fluctuation measurements from Beam Emission Spectroscopy~\cite{van2017}.} reported in \ref{sec:numerical_convergence}. 
Local nonlinear simulations for STEP flat-top operating points on the above grids typically require $5\times 10^5$~CPU-hours on the ARCHER2 high performance computing system; i.e. they are computationally expensive.

\subsection{Flow shear considerations}

\noindent Sensitivity of the linear growth rate to the ballooning parameter $\theta_0 = k_{x}/(\hat{s}k_y)$ is a reliable indicator of a mode's susceptibility to stabilisation by equilibrium flow shear. Linear analysis in \cite{kennedy2023a} shows that the growth rate of the hybrid-KBM instability is highly sensitive to $\theta_0$ and that the mode is stable for all $\theta_0$ above a very small value.  Hybrid-KBM turbulence should therefore be very sensitive to flow shear. 
Lacking any external momentum source from neutral beam injection in STEP, rotation is expected to be modest.  Equilibrium flow shear will be present, however, at the diamagnetic level.   

We will assess the sensitivity of hybrid-KBM turbulence to the equilibrium flow shearing rate $\gamma_E$, and estimate the diamagnetic flow shearing rate using (see e.g. \cite{parisi2020}):
\begin{equation}
\label{eqn:shear}
    \gamma_E^{\mathrm dia} =\frac{1}{B}\biggl(\frac{\partial\Psi}{\partial\rho}\biggr)^2\biggl[\frac{1}{p_{i} n_i e(1+\eta_i)}\biggl(\frac{\partial p_{i}}{\partial \Psi}\Bigr)^2-\frac{1}{n_i e}\frac{\partial^2p_{i}}{\partial\Psi^2}\biggr]\,
\end{equation}
where $\eta_i = L_n/L_{T_i}$ is the ratio of the gradient scale lengths associated with density and ion temperature \textcolor{black}{(which is equal for deuterium and tritium)}, and $p_i$ is the \textcolor{black}{total} ion pressure, \textcolor{black}{and $n_i=n_D+n_T$}. From equation~(\ref{eqn:shear}) the value of $\gamma_E^{\mathrm dia}$ at $\Psi_n = 0.49$ is approximately $\gamma_E^{\mathrm dia} \simeq 0.05\,c_{s}/a$. 
\textcolor{black}{Details on the equilibrium flow shear implementation in CGYRO are reported in \cite{candy2018}.}
Our main purpose is to assess challenges and uncertainties impacting on gyrokinetic calculations of turbulent transport at the STEP reference operating point; one crucial element is clearly to understand the sensitivity of turbulent fluxes to equilibrium flow shear. 

\subsection{Hybrid-KBM turbulence results}
\label{sec:hybridKBMturb}

\noindent We have performed a set of nonlinear simulations using the default grid, for $\gamma_E \in \{0, 0.05, 0.1, 0.2\}\, c_s/a$, where equilibrium flow shear is turned on well into the linear growing phase. 

\textcolor{black}{Previous works (see e.g. \cite{waltz2010,pueschel2013}) show that simulations of  electromagnetic gyrokinetic turbulence do not always reach a saturated state. In finite simulations, however, it can be difficult to distinguish between turbulent fluxes that will never saturate and turbulent fluxes that might eventually saturate potentially at a very large value. In order to more clearly distinguish between simulations with runaway fluxes and those that saturate within the time simulated,  this work introduces a rigorous definition of a saturated state using the augmented Dickey-Fuller test~\cite{dickey1979}; this tests the turbulent fluxes for statistical stationarity.} The null hypothesis of the augmented Dickey-Fuller test is the presence of a unit root, while the alternative hypothesis is the stationarity of the time series. The null hypothesis is rejected when the \texttt{p-value} returned by the statistical test is below a threshold value that is typically chosen between 0.05 and 0.1. In this work, we apply the augmented Dickey-Fuller test to the time trace of the total heat flux and we define the time series to be stationary (i.e. the simulation reaches a robustly steady saturated state) when the \texttt{p-value} is below 0.1. \textcolor{black}{ Every simulation in this paper achieves a saturated state according to this criterion, and wherever time-averaged statistics are reported they are averaged over this saturated state.} 

\begin{figure}
    \centering
    \subfloat[]{\includegraphics[height=0.23\textheight]{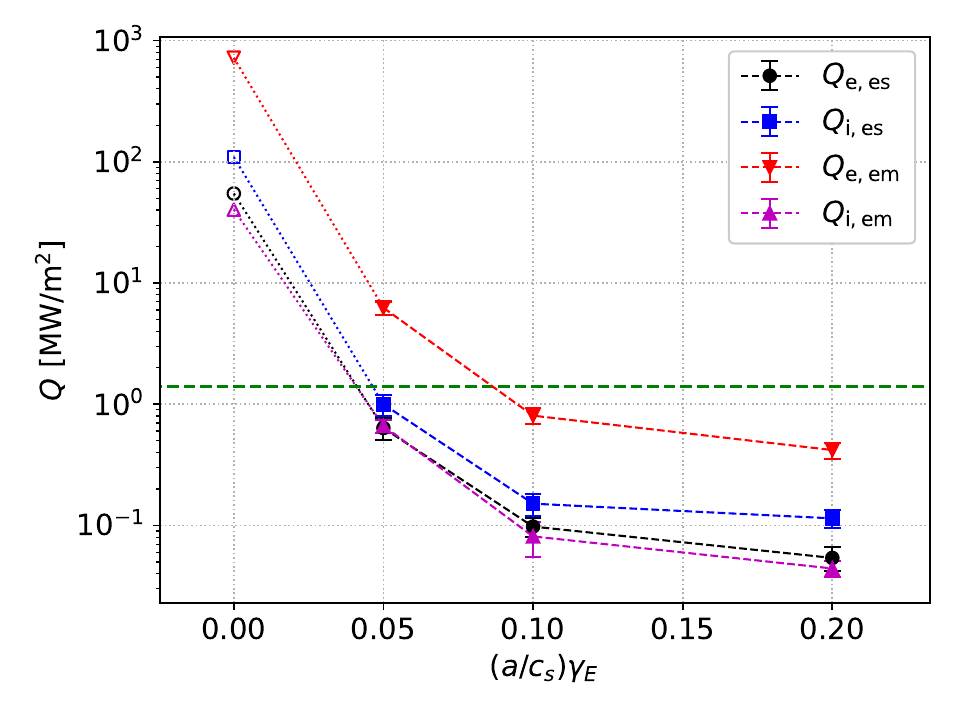}}\,
    \subfloat[]{\includegraphics[height=0.23\textheight]{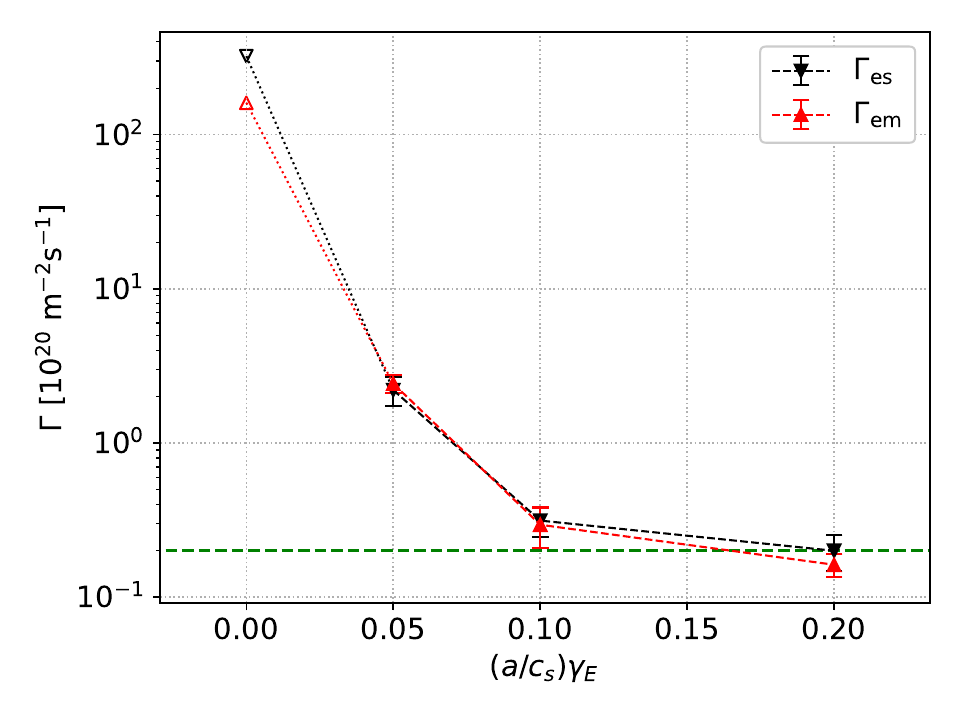}}
    \caption{(a) Electrostatic electron and ion heat fluxes, $Q_\mathrm{e, es}$ and $Q_\mathrm{i, es}$, as well as electromagnetic electron and ion heat fluxes, $Q_\mathrm{e, em}$ and $Q_\mathrm{i, em}$, evaluated from nonlinear simulations at different values of $\gamma_E$. (b) Electrostatic and electromagnetic particle fluxes, $\Gamma_\mathrm{es}$ and $\Gamma_\mathrm{em}$, evaluated from nonlinear simulations at different values of $\gamma_E$. The horizontal dashed lines indicate maximum fluxes that could be consistent with the reference STEP equilibrium, and are taken as the total heat and particle sources, respectively, divided by the area of the $\Psi_n=0.49$ surface. \textcolor{black}{Open markers are used at $\gamma_E=0$ where the turbulent flux bursts (see Figure \ref{fig:timetrace}) influence the time averaged values.}}
    \label{fig:shear_scan}
\end{figure}

Figure~\ref{fig:shear_scan} shows the time averaged heat and particle flux values from nonlinear simulations with various flow shearing rates, $\gamma_E$.
Heat and particle fluxes both decrease with $\gamma_E$.
We note that the electromagnetic electron heat flux, $Q_\mathrm{e, em}$, dominates the total heat flux at all shearing rates considered. The second largest contribution comes from the electrostatic ion heat flux $Q_\mathrm{i, es}$, which is more than a factor of five smaller than $Q_\mathrm{e, em}$. The electrostatic and electromagnetic particle flux values are similar.
At $\gamma_E = 0.05\, c_s/a \sim \gamma_E^{\mathrm dia}$, heat and the particle fluxes exceed the maximum steady-state fluxes that could be provided from the sources assumed to be available in STEP \textcolor{black}{(shown as a horizontal green line in Figure~\ref{fig:shear_scan})}.
Increasing the equilibrium flow shear to $\gamma_E=0.1\, c_s/a$ brings the heat flux below the available source, though the particle flux remains above. We note that, unlike the heat source, fuelling can be more easily increased without compromising fusion gain. If the equilibrium flow shear is increased further the particle flux drops below the maximum available source. 
The same qualitative and quantitative trends of heat and particle fluxes with $\gamma_E$ are also observed in GENE simulations (see \ref{sec:comparison} for a comparison with GENE results).

\subsubsection{Turbulence with \texorpdfstring{$\gamma_E=0$}{TEXT}\\}

\begin{figure}
    \centering
    \subfloat[]{\includegraphics[height=0.22\textheight]{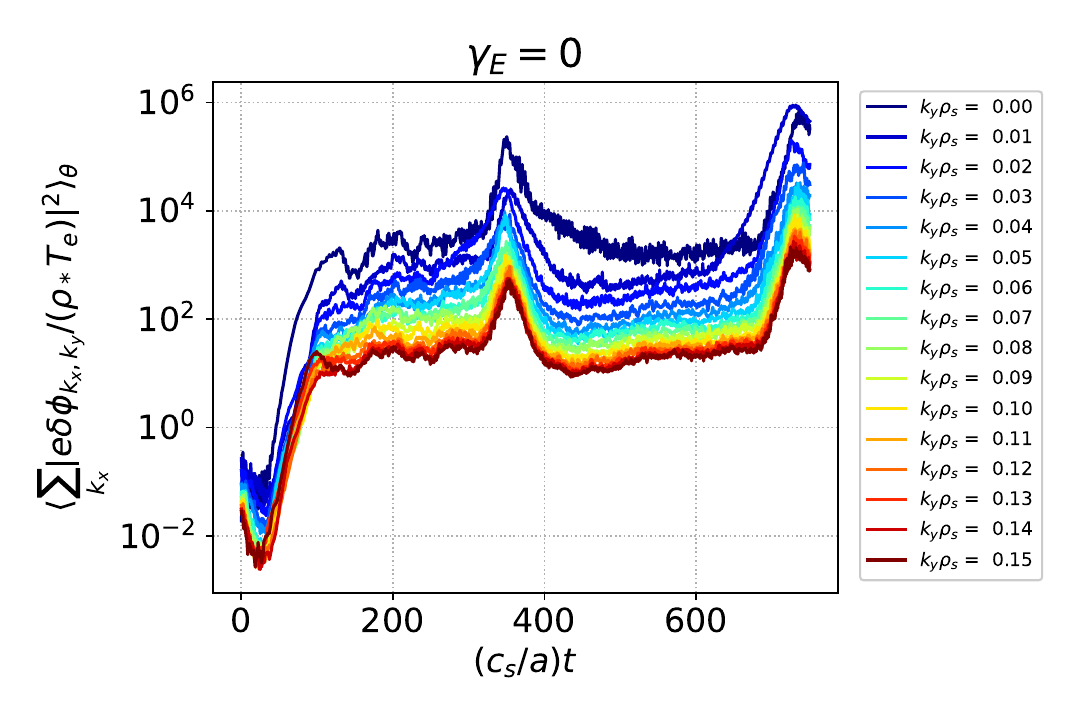}}\,
    \subfloat[]{\includegraphics[height=0.22\textheight]{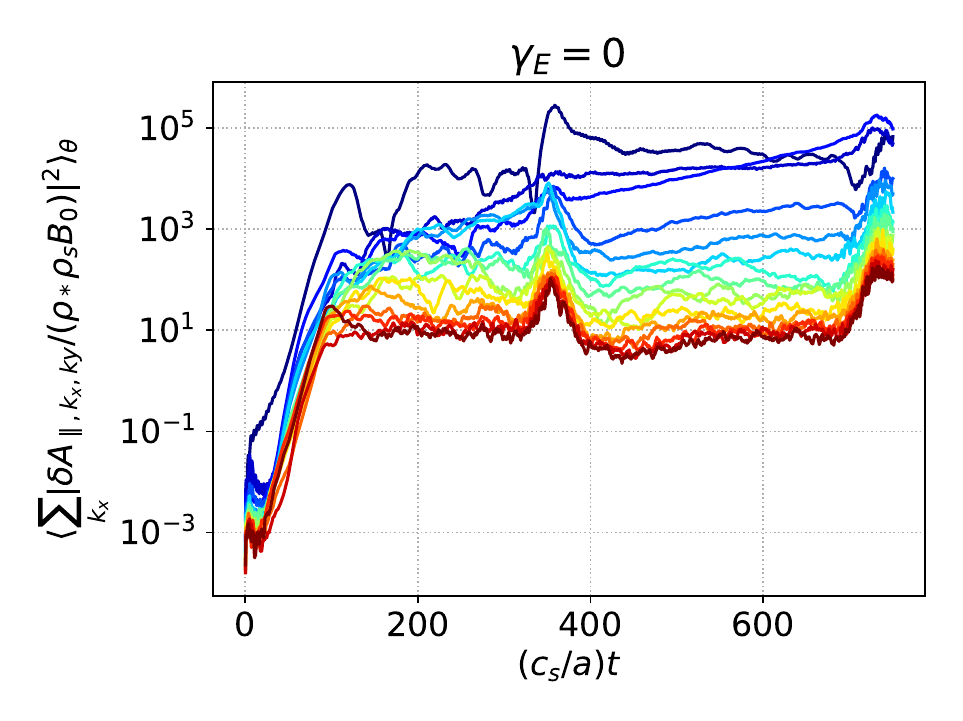}}\\
    \subfloat[]{\includegraphics[height=0.23\textheight]{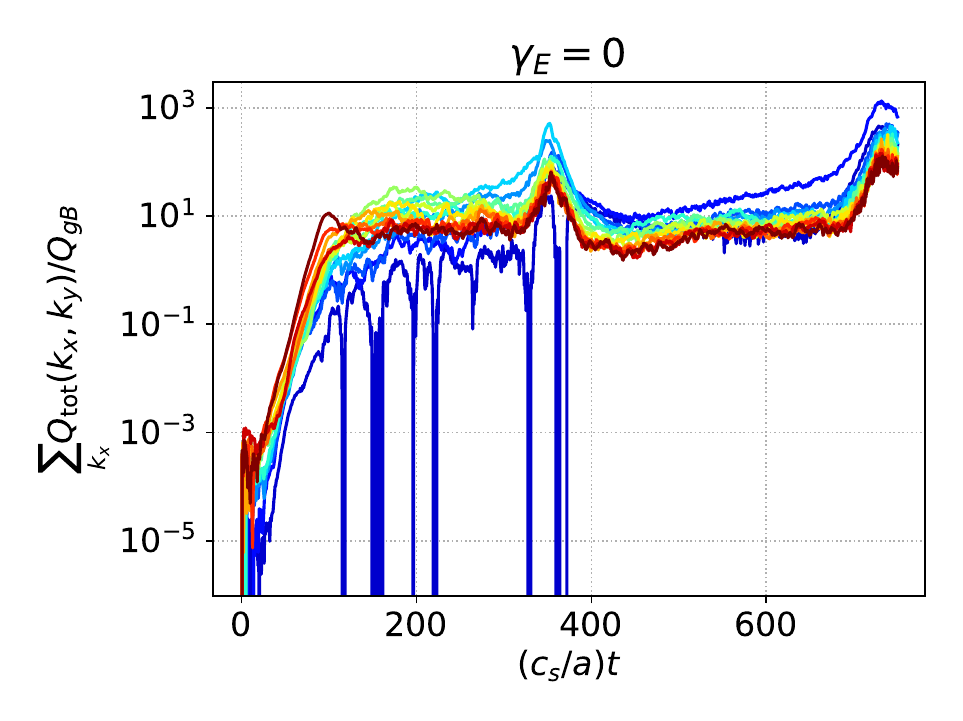}}\,
    \subfloat[]{\includegraphics[height=0.23\textheight]{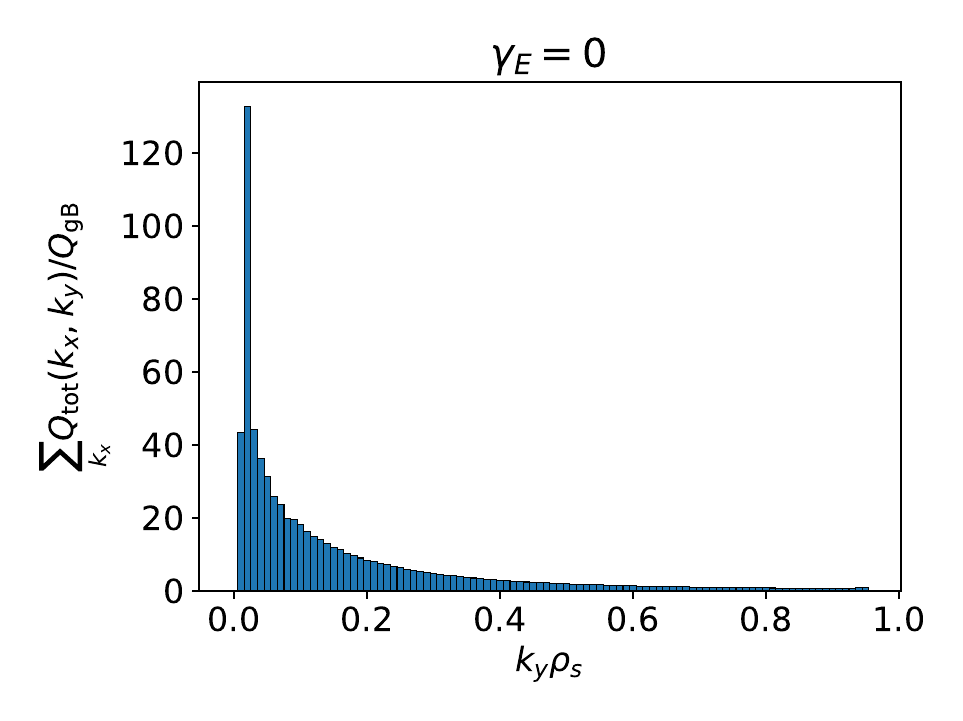}}
    \caption{Time trace of the lowest 16 $k_y$ values of $\delta \phi$ (a), $\delta A_\parallel$ (b) and $Q_\mathrm{tot}$ (c) from the nonlinear simulation without flow shear. (d) $k_y$ spectrum of the total heat flux averaged over time between $t=150\, a/c_s$ and the end of the simulation time.}
    \label{fig:timetrace}
\end{figure}

\begin{figure}
    \centering
    \subfloat[]{\includegraphics[height=0.23\textheight]{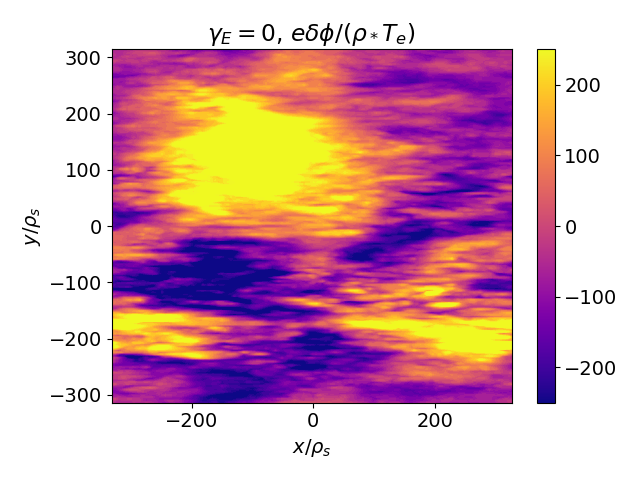}}\,
    \subfloat[]{\includegraphics[height=0.23\textheight]{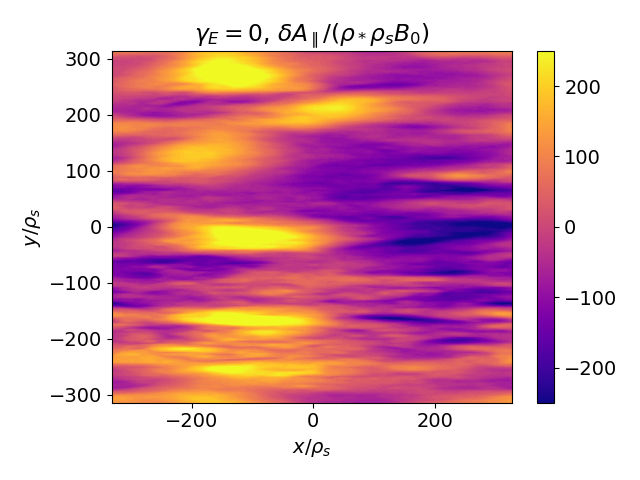}}\\
    \caption{Snapshot contour plots on the outboard midplane (at $\theta=0$) from the final time step  of the simulation without equilibrium flow shear: (a) $\delta \phi$ and (b) $\delta A_\parallel$.}
    \label{fig:real_exb0}
\end{figure}

It is important to first discuss simulations without equilibrium flow shear, before shifting our focus to more physically relevant simulations with $\gamma_{E} > 0$.
Figure~\ref{fig:timetrace} shows time traces of contributions to $\delta \phi$, $\delta A_\parallel$ and $Q_\mathrm{tot}$ \textcolor{black}{(normalised to $Q_\mathrm{gB} = \rho_*^2 c_s n_e T_e\simeq 1.4$~MW/m$^2$ where $\rho_*=\rho_s/a$)} from each of the 16 lowest $k_y$ modes (which  dominate turbulent transport in these simulations) from the nonlinear simulation with $\gamma_E=0$. The lowest $k_y$ modes have the largest amplitudes in $e\delta \phi/(\rho_*T_e)$ and $\delta A_\parallel/(\rho_*\rho_sB_0)$, despite these low $k_y$ modes being only marginally unstable with the lowest $k_y > 0$ mode linearly stable. 
After the initial linear growth phase, the heat flux contributions from various modes exceed 10~$Q_\mathrm{gB}$.
Between $t\simeq 150\,a/c_s$ and $t\simeq 300\,a/c_s$, both fields and fluxes at $k_y\rho_s>0.05$ appear to saturate, though the amplitudes of the $k_y\rho_s\leq0.05$ modes increase weakly. The amplitude of \textcolor{black}{$\delta A_\parallel$} for low $k_y$ modes \textcolor{black}{exceeds} the zonal \textcolor{black}{$\delta A_\parallel$} amplitude at $t\simeq350\,a/c_s$ (see Figure~\ref{fig:timetrace}), which coincides with a first  heat flux burst. \textcolor{black}{After the heat flux burst, the heat flux returns to values similar to the one observed in the time window between $t\simeq 150\,a/c_s$ and $t\simeq 300\,a/c_s$. We note however that $\delta \phi$ and $\delta A_\parallel$ at low $k_y$ modes ($k_y\rho_s < 0.05$) keep increasing slowly in this phase. At $t\simeq 620\,a/c_s$, the $\delta\phi$ and $\delta A_\parallel$ amplitude of the non-zonal low $k_y$ modes exceeds the zonal amplitude, and a second heat flux burst occurs. Apart from the heat flux bursts, the heat flux value remains constant over a quite long time window for $t>150\,a/c_s$. The augmented Dickey-Fuller test applied to the heat flux over  $t>150\,a/c_s$ return a \texttt{p-value} of 0.01, thus supporting the stationarity of the heat flux time trace. We note that continuing this simulation further is extremely expensive due to the very small time step and it does not change the main outcome: turbulent fluxes are three orders of magnitude larger than values compatible with the available STEP sources in the absence of sheared flows. (This conclusion is also supported by similar results from  GENE simulations that are reported in \ref{sec:comparison}).} 
Figure~\ref{fig:timetrace}~(d) shows the $k_y$ spectrum of the heat flux averaged over time from \textcolor{black}{$t=150\,a/c_s$} to the end of the simulation. The heat flux spectrum peaks at $k_y\rho_s\simeq 0.02$ and decays at large $k_y$.  
Figure~\ref{fig:real_exb0} shows snapshot contour plots of $\delta \phi$ and $\delta A_\parallel$ at the outboard mid-plane ($\theta=0$), taken from the last time step. The fluctuation amplitudes of $e\delta\phi/(\rho_*T_e)$ and $\delta A_\parallel/(\rho_*\rho_sB_0)$ are comparable, demonstrating the strongly electromagnetic nature of the turbulence \footnote{We note that although $\delta B_{\parallel}$ is essential for hybrid-KBM linear instability, $\delta B_\parallel/(\rho_* B_0)$ is substantially smaller than $e\delta\phi/(\rho_*T_e)$ and $\delta A_\parallel/(\rho_*\rho_sB_0)$ in all the nonlinear simulations performed in this work.}.
The turbulent structures are highly elongated radially, extending to $\sim 100\,\rho_s$, which corresponds here to approximately 50~cm.  
Similar large flux states have been found using various radial box sizes and radial resolutions as well as with different GK codes (see \ref{sec:numerical_convergence} for details), suggesting that for this local equilibrium large fluxes are a robust prediction of local gyrokinetic simulations. Given the large radial extent of the turbulent structures, however, global effects not captured in the local approach may also affect saturation; \textcolor{black}{e.g. background equilibrium profile variation effects, or another recently proposed global  mechanism whereby global zonal flows are enhanced by the radial transfer of entropy \cite{masui2022}.} Global gyrokinetic simulations \emph{that include $\delta B_\parallel$} may more accurately predict turbulent fluxes in this regime, and constitute a possible future research direction. 
We note, however, that local nonlinear simulations with modest flow shear produce less radially extended structures (e.g. see snapshots of $\delta \phi$ and $\delta A_\parallel$ in Figure~\ref{fig:real_exb005}), and that further actuators and mechanisms (see \secref{sec: sensitivity to local parameters}) also result in turbulent states with lower transport that are more amenable to local gyrokinetics.  

\subsubsection{Turbulence with finite \texorpdfstring{$\gamma_E$}{TEXT}\\}

\begin{figure}
    \centering
    \subfloat[]{\includegraphics[height=0.22\textheight]{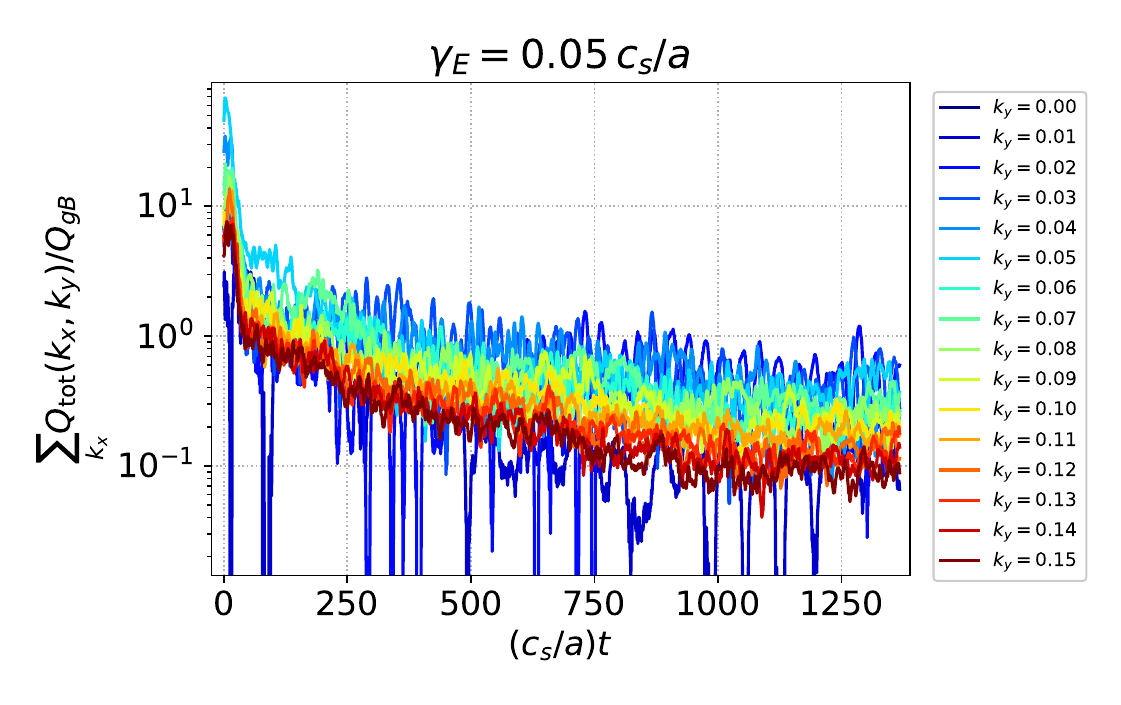}}\,
    \subfloat[]{\includegraphics[height=0.22\textheight]{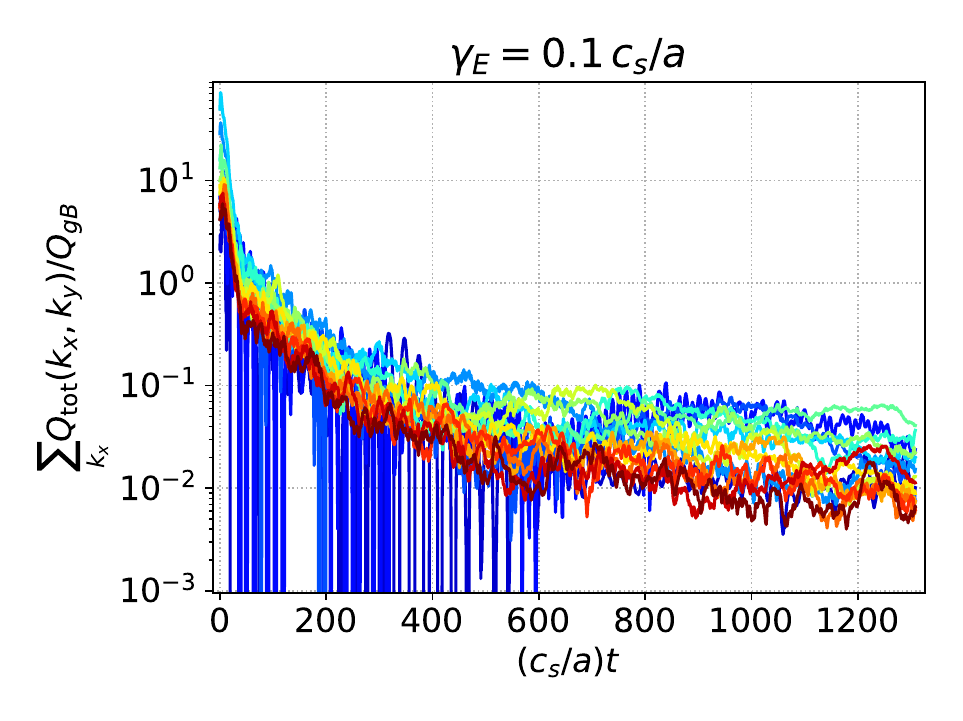}}\\
    \subfloat[]{\includegraphics[height=0.24\textheight]{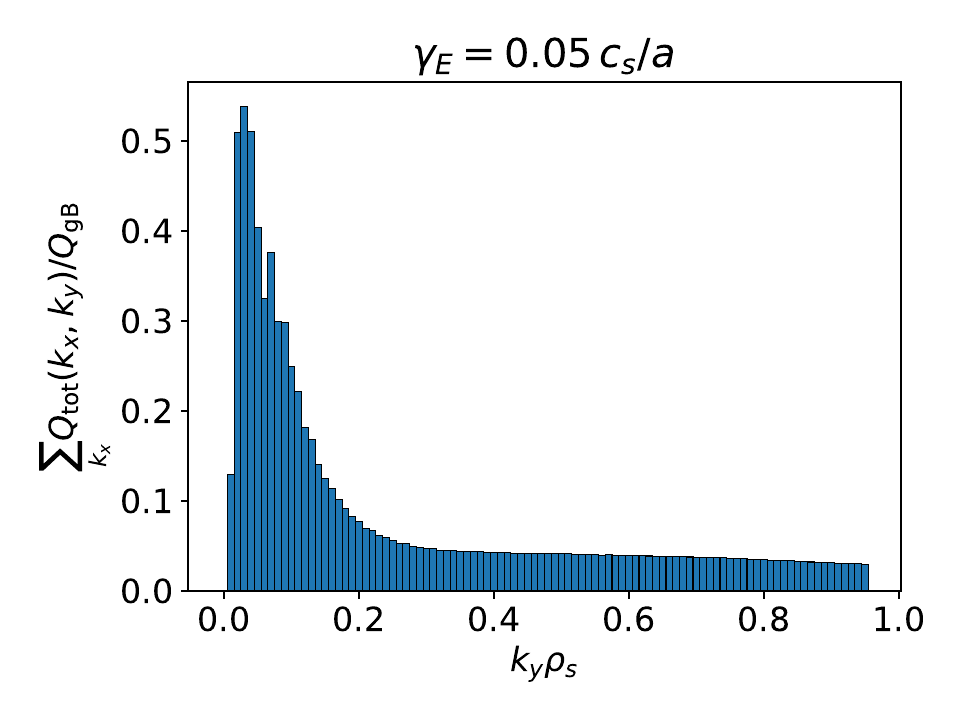}}\,
    \subfloat[]{\includegraphics[height=0.24\textheight]{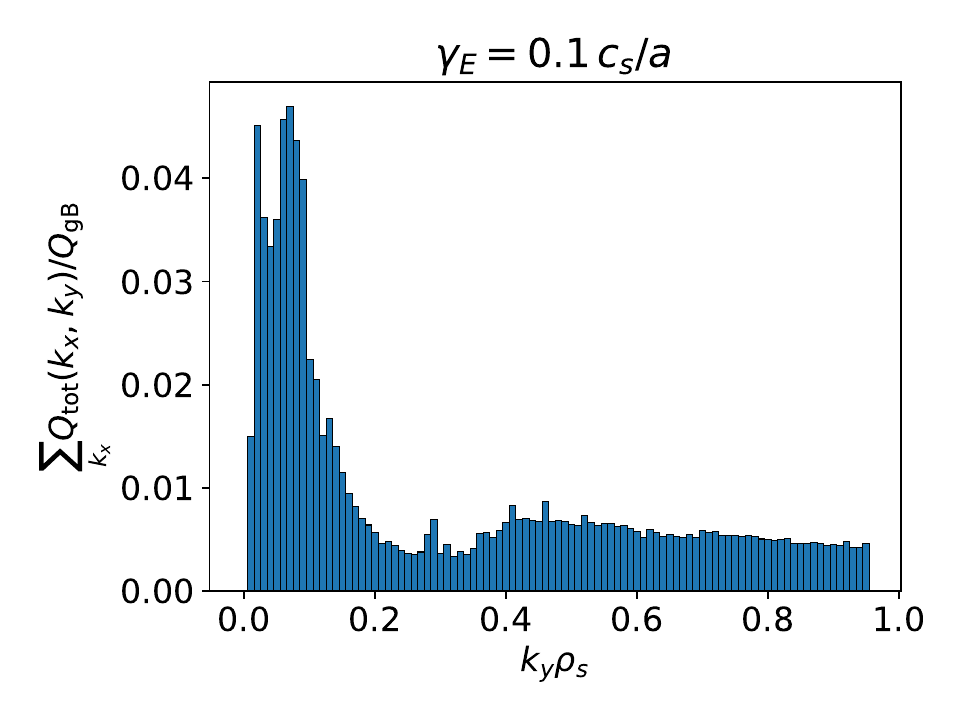}}
    \caption{Time trace of the first 16 $k_y$ modes (excluding $k_y=0$) of the heat flux from the nonlinear simulation with $\gamma_E=0.05\ c_s/a$ (a) and $\gamma_E = 0.1\, c_s/a$ (b). The embedded figures in (a) and (b) show a zoom over the last 300 ion sound times of the heat flux time trace, also including the total heat flux summed over $k_y$ (black line). Heat flux as a function of $k_y$ from the nonlinear simulation with $\gamma_E=0.05\ c_s/a$ (c) and $\gamma_E = 0.1\, c_s/a$ (d).}
    \label{fig:timetrace_shear}
\end{figure}

We now focus on the simulations with finite equilibrium flow shear. Figure~\ref{fig:timetrace_shear} shows contributions to the heat flux from the lowest 16 $k_y$ modes as functions of time \textcolor{black}{for the simulations with $\gamma_E=0.05\,c_s/a$ and $\gamma_E=0.1\,c_s/a$}, where equilibrium flow shear is turned on at $t=0$ with the initial condition taken from after the linear growth phase in a simulation with $\gamma_E=0$.
The heat flux decreases in the presence of equilibrium flow shear, with a sharp initial reduction followed by a phase of weaker decay. The two simulations take until approximately $t\simeq 1000\,a/c_s$  to reach a saturated state. The $k_y$ spectrum of the time averaged heat flux in the saturated phase is shown in Figure~\ref{fig:timetrace_shear}~(c)~and~(d).
\textcolor{black}{At $\gamma_E=0.05\,c_s/a$,} the total heat flux peaks around $k_y\rho_s = 0.03$  and decays for $k_y\rho_s>0.1$ to a small tail contribution to the flux from high $k_y \rho_s$. 
\textcolor{black}{At $\gamma_E=0.1\,c_s/a$, the spectrum is broader, with the main peak shifting to higher $k_y$ and a second distinct lower amplitude peak at $k_y\rho_s\simeq 0.45$.}
In both cases, low $k_y$ modes make a relatively large contribution to the heat flux, although the peak of the spectrum is broader and at larger $k_y$ compared to Figure~\ref{fig:timetrace}~(d) for $\gamma_E=0$.
Equilibrium flow shear clearly acts to suppress the formation of large amplitude radially extended turbulent structures.

\begin{figure}
    \centering
    \subfloat[]{\includegraphics[height=0.23\textheight]{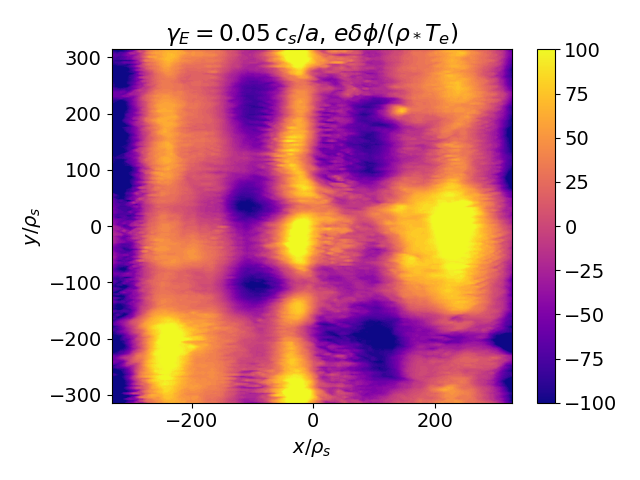}}\,
    \subfloat[]{\includegraphics[height=0.23\textheight]{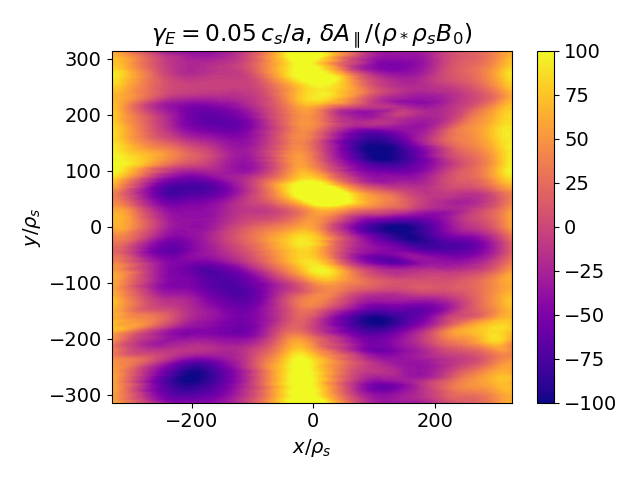}}
    \caption{Snapshot contour plots on the outboard midplane (at $\theta=0$) from the final time step  of the simulation with $\gamma_E=0.05\,c_s/a$: (a) $\delta \phi$, and (b) $\delta A_\parallel$.}
    \label{fig:real_exb005}
\end{figure}

\begin{figure}
    \centering
    \subfloat[]{\includegraphics[height=0.23\textheight]{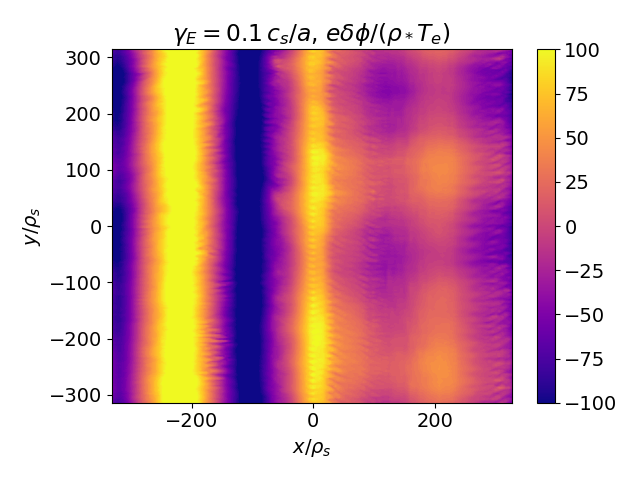}}\,
    \subfloat[]{\includegraphics[height=0.23\textheight]{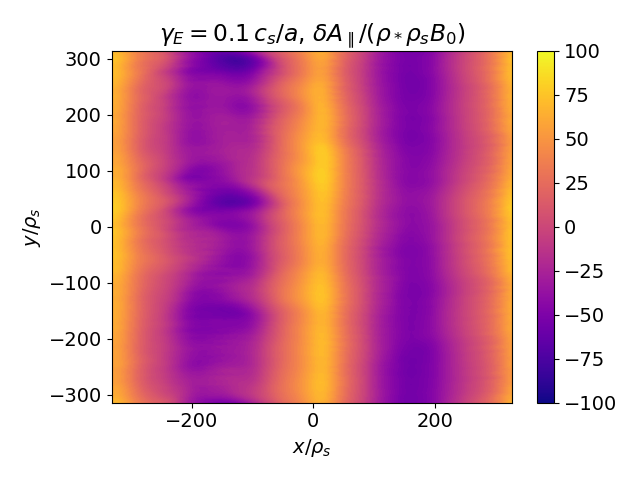}}
    \caption{Snapshot contour plots on the outboard midplane (at $\theta=0$) from the final time step  of the simulation with $\gamma_E=0.1\,c_s/a$: (a) $\delta \phi$, and (b) $\delta A_\parallel$.}
    \label{fig:real_exb01}
\end{figure}

The simulations with finite $\gamma_E$ are characterised by relatively strong zonal flows ($\sum_{k_x} \delta \phi(k_x, k_y=0)$) and zonal fields ($\sum_{k_x} \delta A_\parallel(k_x, k_y=0)$), which reduce considerably the radial extent of the turbulent structures, as shown in Figure~\ref{fig:real_exb005}~and~\ref{fig:real_exb01}. The amplitudes of $e\delta \phi/(\rho_*T_e)$ and $\delta A_\parallel/(\rho_*\rho_s B_0)$ are again comparable, as expected from electromagnetic turbulence.
We highlight that the  $\delta A_\parallel/(\rho_*\rho_s B_0)$ amplitude is smaller at $\gamma_E\simeq 0.1\,c_s/a$ than at $\gamma_E \simeq 0.05\,c_s/a$. In addition, the radial extents of structures in $\delta \phi$ and $\delta A_\parallel$ at $\gamma_E\simeq 0.1\,c_s/a$ are significantly smaller at $\sim 30\rho_s$ (more easily visible if the zonal component is removed).

\textcolor{black}{The analysis presented in this section suggests that, while equilibrium flow shear reduces the turbulent fluxes, at the diamagnetic level it is insufficient to reduce the hybrid-KBM heat and particle fluxes to levels compatible with the assumed sources on the chosen surface from the STEP operating point. 
This motivates an assessment of the sensitivity of hybrid-KBM turbulent transport to key local equilibrium parameters, to seek routes for optimising transport at the STEP flat-top operating point.}  

\section{Sensitivity studies and mitigation mechanisms}
\label{sec: sensitivity to local parameters}

This section is divided in two parts. In the first part, we assess the sensitivity of turbulent fluxes to the pressure gradient that drives the hybrid-KBM instability in nonlinear simulations with equilibrium flow shear. 
In the second part, we explore additional possible mechanisms other than equilibrium flow shear to mitigate the impact of hybrid-KBM. This includes some sensitivity scans to $\beta_{e}$ and $q$, with the aim of identifying more favorable regimes for a STEP flat-top operating point. 

\subsection{Sensitivity on local pressure gradient in simulations with \texorpdfstring{$\gamma_E=0.05\,c_s/a$}{TEXT}}

Section \secref{sec:hybridKBMturb} shows that turbulent fluxes can exceed the target value for this STEP flat-top operating point even in the presence of equilibrium flow shear. Here we explore the sensitivity of hybrid-KBM fluxes to the pressure gradient drive to assess whether target transport fluxes can be achieved through a modest change to the pressure gradient.

We perform nonlinear simulations with three different values of the equilibrium pressure gradient, $L_{p,\mathrm{ref}}/L_p\in \{0.8, 1, 1.2\}$, where $a/L_{p,\mathrm{ref}}$ denotes the reference value of the local pressure gradient. In this scan the pressure gradient is varied by scaling both the \textcolor{black}{electron and ion} density and temperature gradients simultaneously by the same factor. 
All other local equilibrium parameters are kept constant in this scan, except for $\beta^\prime\equiv \beta \partial \ln p/\partial \rho$, which is scaled consistently with the pressure gradient to retain the important effect of $\beta^\prime$ on the linear hybrid-KBM instability~\cite{kennedy2023a}.
All simulations include equilibrium flow shear with a constant value of $\gamma_E=0.05\,c_s/a$.
The numerical resolution of the simulation at $a/L_p = 1.2(a/L_p)_\mathrm{ref}$ is identical to that used in the nominal case, while the simulation at $a/L_p = 0.8(a/L_p)_\mathrm{ref}$ requires a higher $k_{y,\mathrm{max}}\rho_s=1.28$  to resolve the linear spectrum (discussed later). 

\begin{figure}
    \centering
    \subfloat[]{\includegraphics[width=0.48\textwidth]{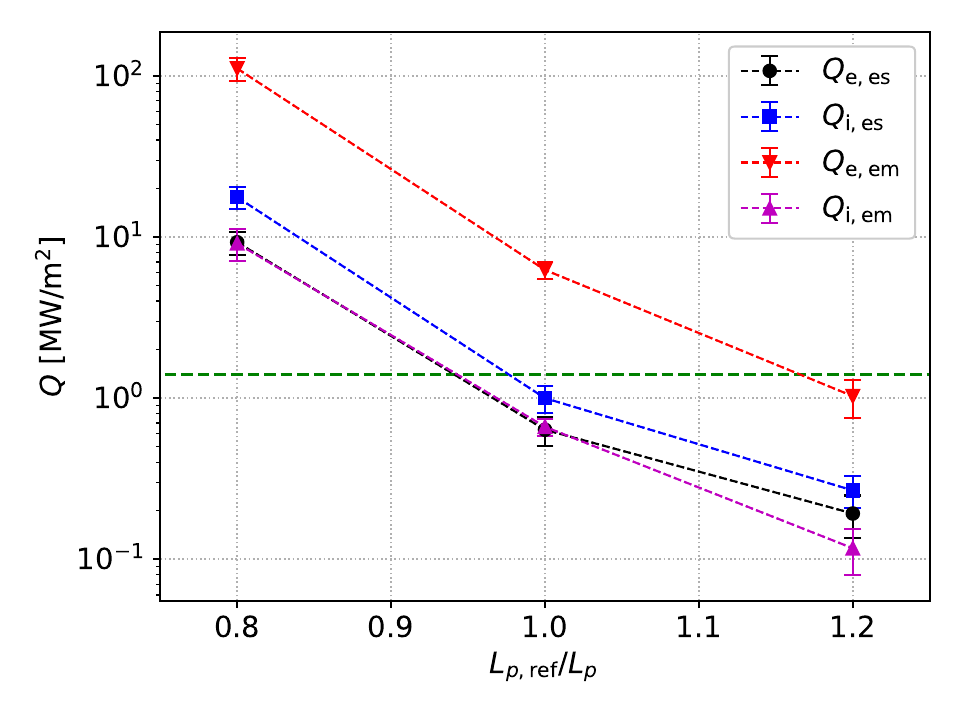}}\,
    \subfloat[]{\includegraphics[width=0.48\textwidth]{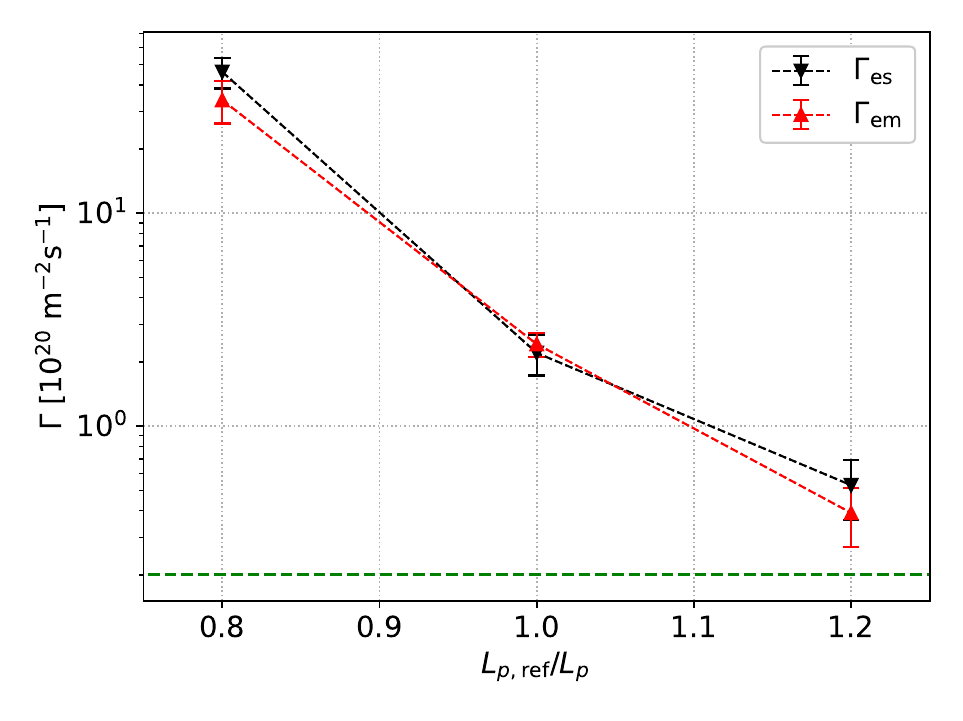}}
    \caption{Electrostatic and electromagnetic electron and ion heat (a) and particle (b) fluxes from nonlinear simulations with different pressure gradient values where $\beta'$ is varied consistently with $L_{p,\mathrm{ref}}/L_p$, \textcolor{black}{where $a/L_{p,\mathrm{ref}}$ denotes the reference value of the total equilibrium pressure gradient.} The horizontal dashed line indicates the target value.}
    \label{fig:alp_scan}
\end{figure}

Figure~\ref{fig:alp_scan} shows the saturated heat and particle fluxes for the different values of $a/L_p$.  
We observe a strong dependence of turbulent fluxes on the pressure gradient. The saturated heat flux at $L_{p,\mathrm{ref}}/L_p = 1.2$ drops below the target flux. 
Whilst the electron electromagnetic heat flux dominates at all pressure gradients, the relative contribution from the electrostatic channel is significantly higher at $L_{p,\mathrm{ref}}/L_p = 1.2$ compared to the other two cases.
This sensitivity study to pressure gradient (and $\beta'$) highlights the beneficial stabilizing effect of increasing $\beta'$ on hybrid-KBM turbulent transport. Furthermore we note that the beneficial effect on transport at higher $a/L_p$ would be further reinforced if equilibrium flow shear were scaled proportionally with the ion pressure gradient as would be expected for diamagnetic flows, but $\gamma_E$ was held fixed in this scan. Decreasing the pressure gradient has the opposite effect and leads to much higher turbulent fluxes.  Local equilibrium scans, like the $a/L_p$ scan presented here, reveal valuable insights, but in a truly consistent scan the whole global MHD equilibrium would evolve such that many local equilibrium parameters, like $q$, $\hat{s}$ and $\beta$ (and potentially the turbulent transport regime) might also change significantly.    

\begin{figure}
    \centering
    \subfloat[]{\includegraphics[width=0.48\textwidth]{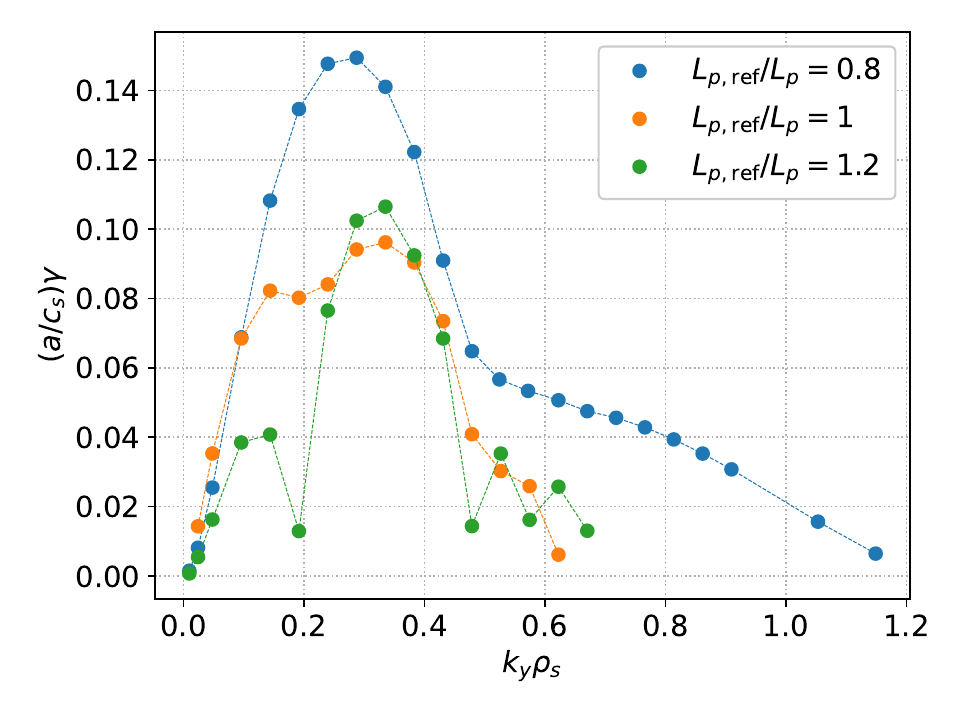}}\,
    \subfloat[]{\includegraphics[width=0.48\textwidth]{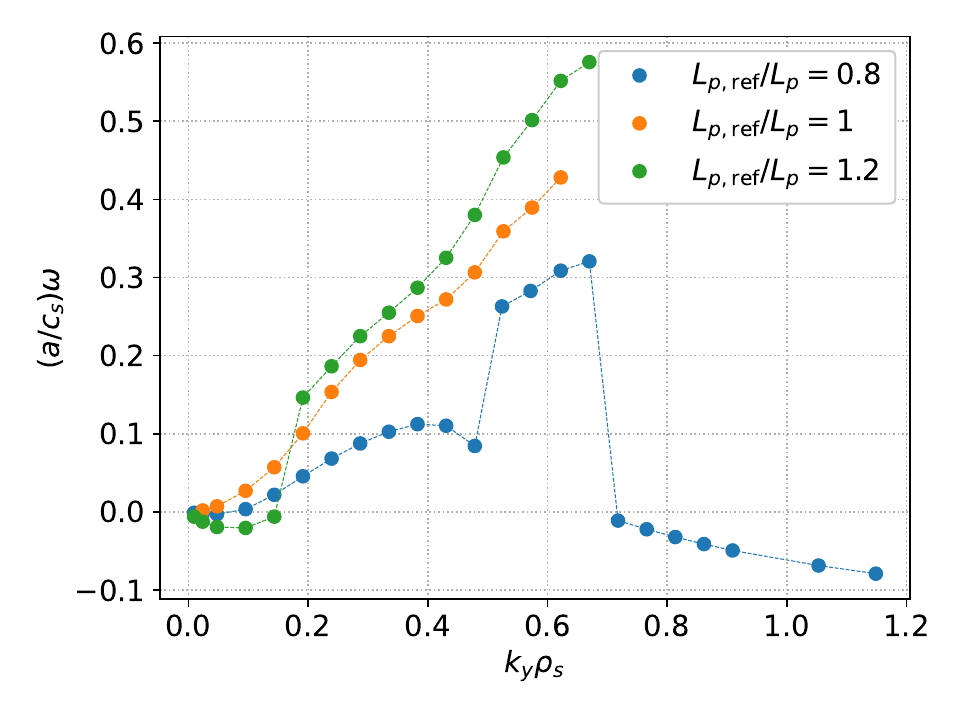}}
    \caption{Growth rate (a) and mode frequency (b) from linear simulations with three different values of local pressure gradient. The value of $\beta^\prime$ is varied consistently with the pressure gradient.}
    \label{fig:linear_alp}
\end{figure}

Figure~\ref{fig:alp_scan} is consistent with the strong, and somewhat counter-intuitive, dependence of the hybrid-KBM linear growth rate on $a/L_p$ (and $\beta^{\prime}$) that is illustrated in Figure~\ref{fig:linear_alp}.
\textcolor{black}{The linear analysis of \cite{kennedy2023a} shows that the hybrid-KBM growth rate increases with density and temperature gradients when all the other parameters are held constant. On the other hand, we find here that the boost to the linear drive from the pressure gradient increase is more than compensated by increased stabilisation from the impact on local equilibrium geometry of higher $\beta^\prime$ (e.g. through the modification of magnetic drifts \cite{roach1995,bourdelle2003}).} 
The growth rates at $L_{p,\mathrm{ref}}/L_p = 0.8$ are larger than in the nominal case, except at low $k_y$ where they are comparable, while the growth rates in the $L_{p,\mathrm{ref}}/L_p = 1.2$ case are smaller than in the nominal case for $k_y\rho_s<0.3$, and they are comparable for $k_y\rho_s\geq 0.3$. The mode frequency values of the nominal and larger pressure gradient cases are comparable.

\subsection{Possible mechanisms to mitigate hybrid-KBMs}
\label{sec:mitigation}

We explore here the impact of local equilibrium parameters, $\beta_e$ and $q$, on hybrid-KBM transport fluxes.  Unless it is explicitly stated otherwise, only the scan parameter is varied and other local parameters are held constant to isolate the effect of each parameter and shed light on physical mechanisms that may mitigate the turbulent transport. In all the scans in this section we set $\gamma_E=0$.

\subsubsection{Sensitivity to \texorpdfstring{$\beta_e$}{TEXT}\\}  
\label{sec:beta}
We focus first on the effect of $\beta$. Nonlinear simulations at different values of $\beta_{e}$ are performed varying $\beta'=\beta_e/L_p$ consistently (with the other local parameters kept constant). 
Numerical resolutions are chosen to properly resolve the linear spectrum of the dominant instability\footnote{In particular, the simulations with $\beta_e=0.02$ and $\beta_e=0.005$ are carried out with $n_{k_y}=128$ and $n_{k_x}=64$, while the simulation with $\beta_e=0.16$ is carried out with $n_{k_y}=32$ and $n_{k_x}=256$.}. 

\begin{figure}
    \centering
    \subfloat[Heat flux]{\includegraphics[width=0.48\textwidth]{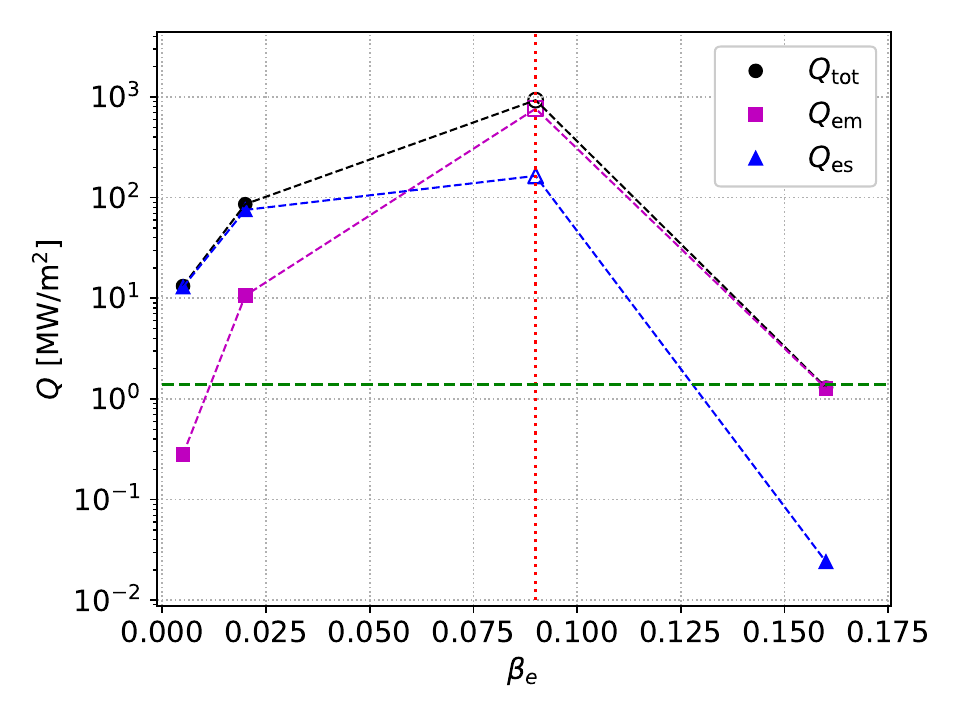}}\,
    \subfloat[Particle flux]{\includegraphics[width=0.48\textwidth]{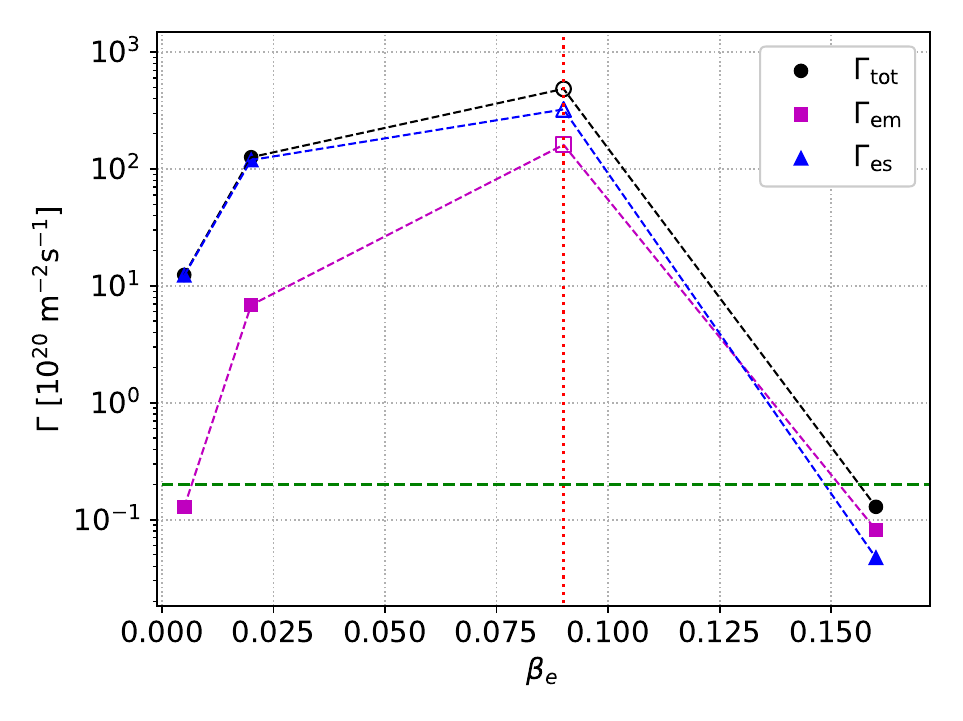}}
    \caption{Total heat (a) and particle (b) fluxes from CGYRO nonlinear simulations at various values of $\beta_e$ with $\beta'$ varying consistently. The dashed vertical line denotes the reference value of $\beta_e$. \textcolor{black}{Open markers are used for the reference simulation where the flux bursts influence the time averaged values.}  } 
    \label{fig:beta}
\end{figure}

\begin{figure}
    \centering
    \subfloat[]{\includegraphics[width=0.48\textwidth]{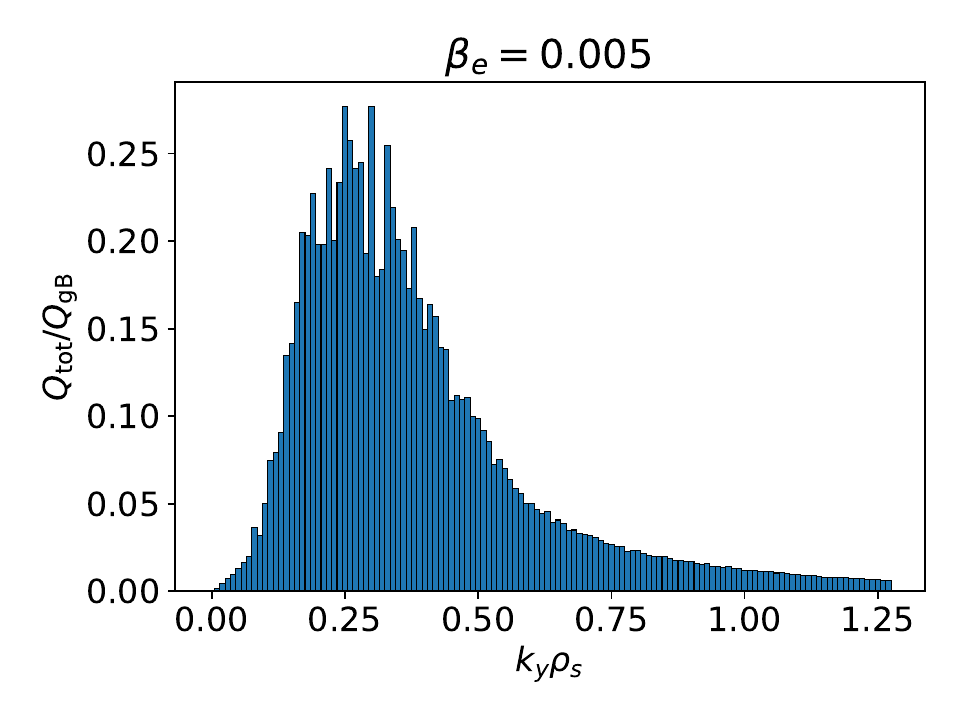}}\,
    \subfloat[]{\includegraphics[width=0.48\textwidth]{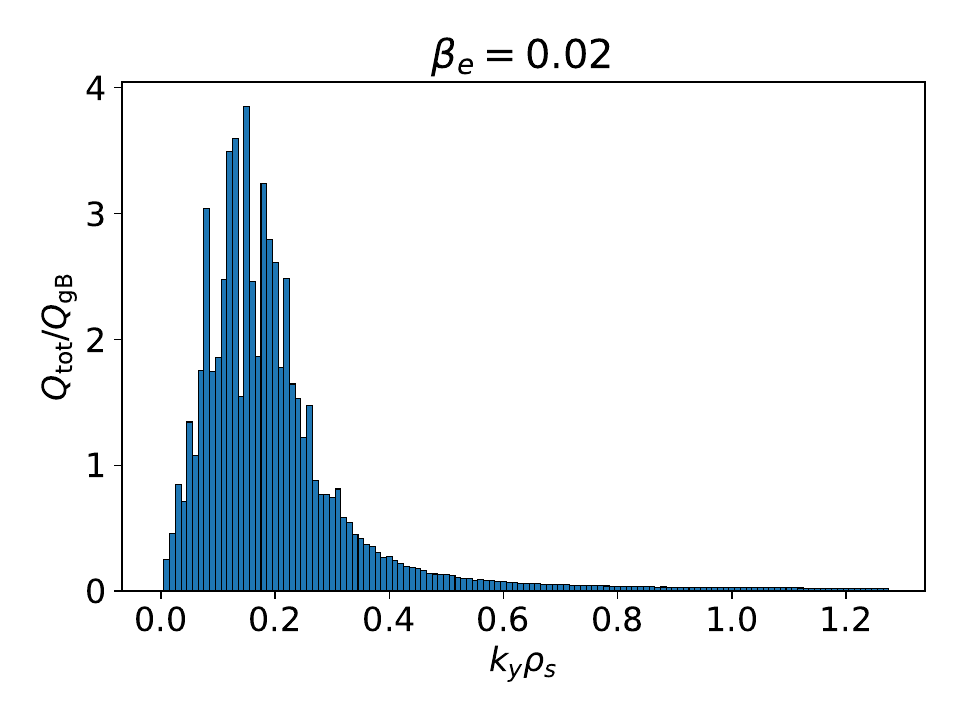}}
    \caption{Total heat flux $k_y$ spectrum from the nonlinear simulation with $\beta_e=0.005$ (a) and $\beta_e=0.02$ (b).}
    \label{fig:beta_ky}
\end{figure}

Figure~\ref{fig:beta} shows the total heat and particle fluxes from nonlinear simulations at different values of $\beta_{e}$.
At $\beta_{e} < 0.025$, the particle and heat transport is mainly electrostatic and is driven by an ITG/TEM instability. As pointed out by \cite{kennedy2023a}, the hybrid-KBM couples to an ITG/TEM instability at low $\beta$. The electrostatic heat and particle fluxes remain above 100~MW/m$^2$ and 10$^{22}$ particles/(m$^2$s) even at $\beta_{e} = \beta_{e,\mathrm{ref}}/4$, and presumably reduced $\beta^{\prime}$ stabilisation plays a significant role in maintaining such large turbulent fluxes. 
In contrast to the low $k_y$ modes at nominal $\beta$, the amplitude of  $\delta A_\parallel/(\rho_*\rho_sB_0)$ is smaller than the $e\delta \phi/(\rho_* T_e)$ one at lower $\beta$, which is consistent with a smaller electromagnetic drive.
In particular, Figure~\ref{fig:beta_ky}  shows that the heat flux contribution from $k_y\rho_s\leq 0.1$ is significantly smaller than the contribution from $k_y\rho_s>0.1$ at lower $\beta_e$, especially at $\beta_e=0.005$.

At $\beta_e=0.16$, above the nominal value, electromagnetic contributions dominate transport and the total heat flux drops considerably from its nominal value to $Q_\mathrm{tot}\simeq 2$~MW/m$^2$, which is close to the assumed transport flux for this STEP flat-top operating point. 
This is a consequence of increased $\beta'$ stabilisation of the hybrid-KBM, which affects all the unstable $k_y$ modes. 
The turbulent transport is dominated by MTMs, which emerge as the dominant linear instability at large $\beta_e$, as detailed in \cite{kennedy2023a}.  This local equilibrium at $\beta_e=0.16$ has considerably more favourable transport than the reference case at $\beta_e=0.09$, if it could be accessed through a route passing through lower $\beta$ that avoided prohibitive transport losses from hybrid-KBMs. However, as noted in \cite{kennedy2023a}, a global equilibrium with this local $\beta_e$ would likely exceed the limiting $\beta$ for effective control of Resistive Wall Modes.

\begin{figure}
    \centering
    \subfloat[]{\includegraphics[width=0.48\textwidth]{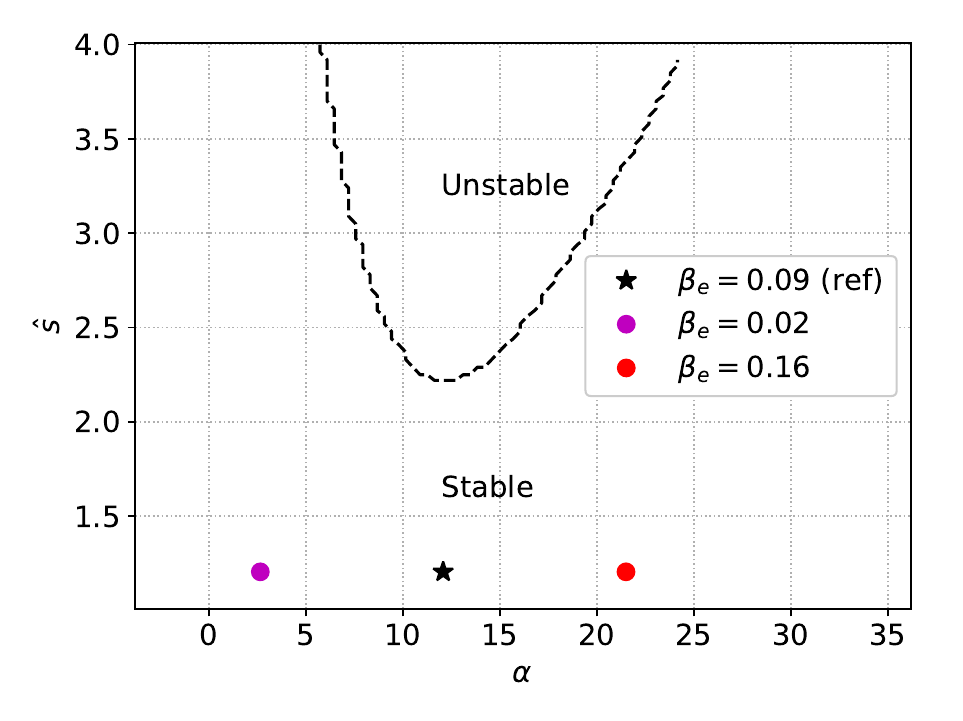}}\,
    \subfloat[]{\includegraphics[width=0.48\textwidth]{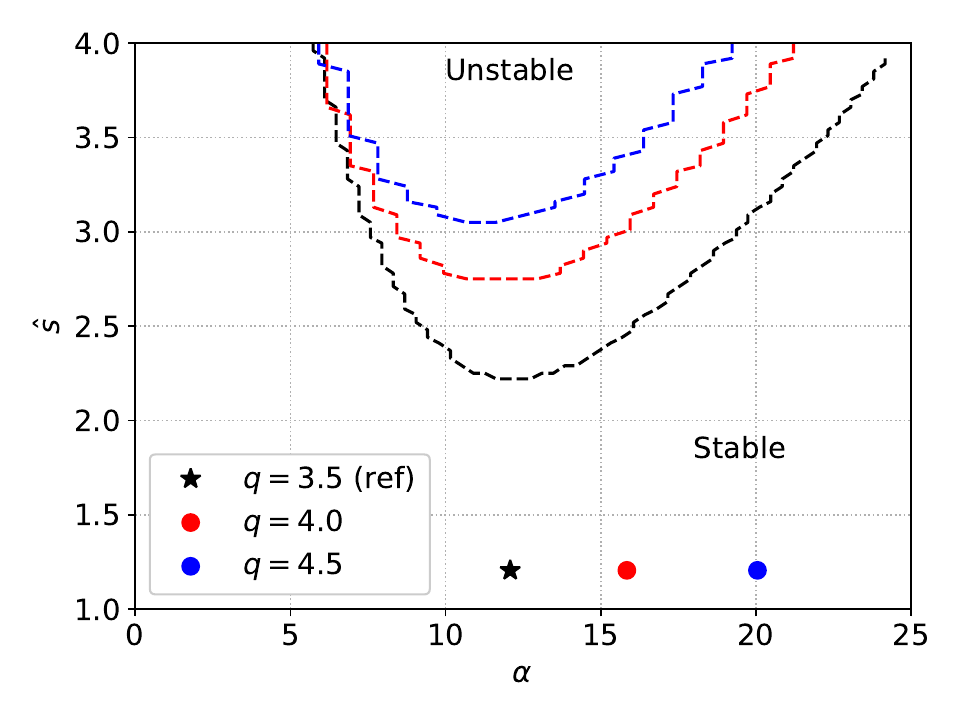}}
    \caption{Ideal ballooning stability boundary in the $\hat{s}-\alpha$ plane of STEP-EC-HD. Panel (a) shows the position in the $\hat{s}-\alpha$ plane of the reference equilibrium and two cases with different $\beta_e$. Panel (b) shows the ideal ballooning boundary for $q=3.5$, $q=4$ and $q=4.5$ cases.}
    \label{fig:salpha}
\end{figure}

Figure~\ref{fig:salpha} suggests that the mitigation of hybrid-KBMs as $\beta_e$ may be related to the relative distance between the local equilibrium point on an  $\hat{s}-\alpha$ diagram and the ideal $n=\infty$ ballooning boundary (where $\alpha = -Rq^2\beta^\prime$). The equilibrium points at $\beta_e=0.02$ and $\beta_e=0.16$ are further from the stability boundary than the reference equilibrium at $\beta_e=0.09$, potentially indicating a reduced electromagnetic contribution to the hybrid-KBM drive, and a subsequent reduction of  heat and particle fluxes driven by the this component of the turbulence.

\subsubsection{Sensitivity to $q$\\} \label{sec:q}
\begin{figure}
    \centering
    \subfloat[]{\includegraphics[width=0.48\textwidth]{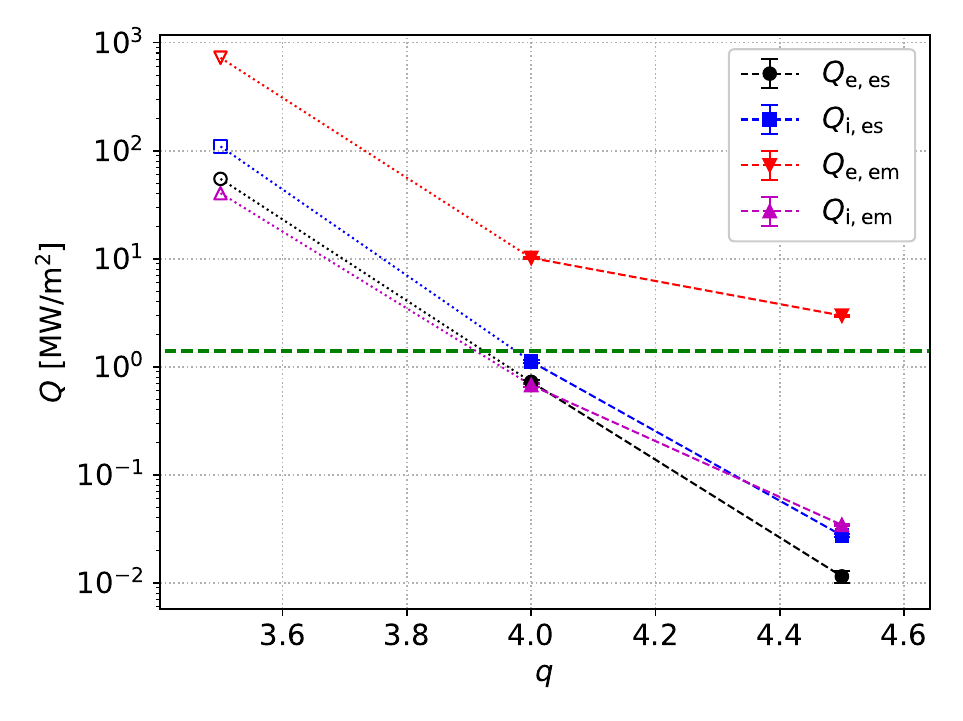}}\,
    \subfloat[]{\includegraphics[width=0.48\textwidth]{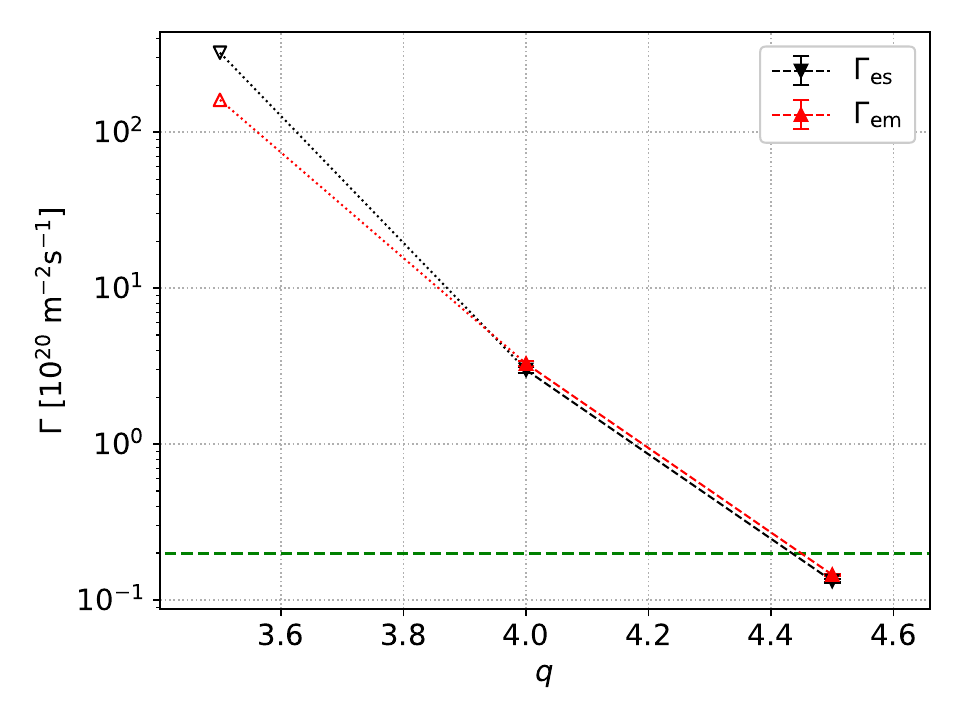}}
    \caption{Electrostatic and electromagnetic electron and ion heat (a) and particle (b) fluxes from nonlinear simulations with different safety factor values. The horizontal dashed line denotes the value of the heat and particle fluxes at  evaluated from the available assumed STEP sources.  \textcolor{black}{Open markers are used for the reference simulation where the flux bursts influence the time averaged values.} } 
    \label{fig:sq_scan}
\end{figure}

\begin{figure}
    \centering
    \subfloat[]{\includegraphics[width=0.48\textwidth]{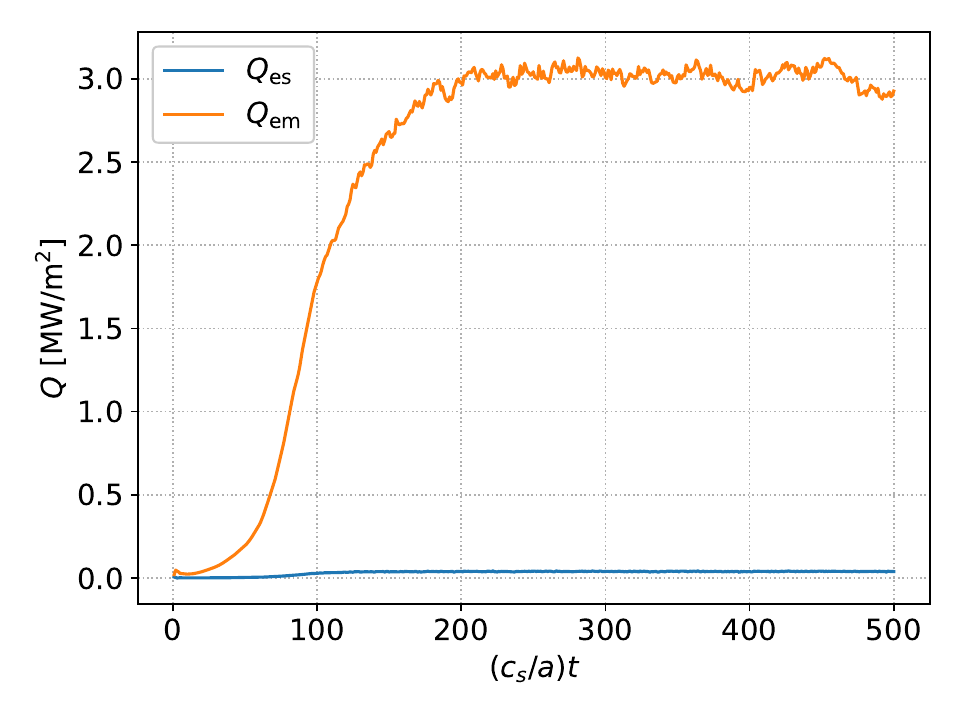}}\,
    \subfloat[]{\includegraphics[width=0.48\textwidth]{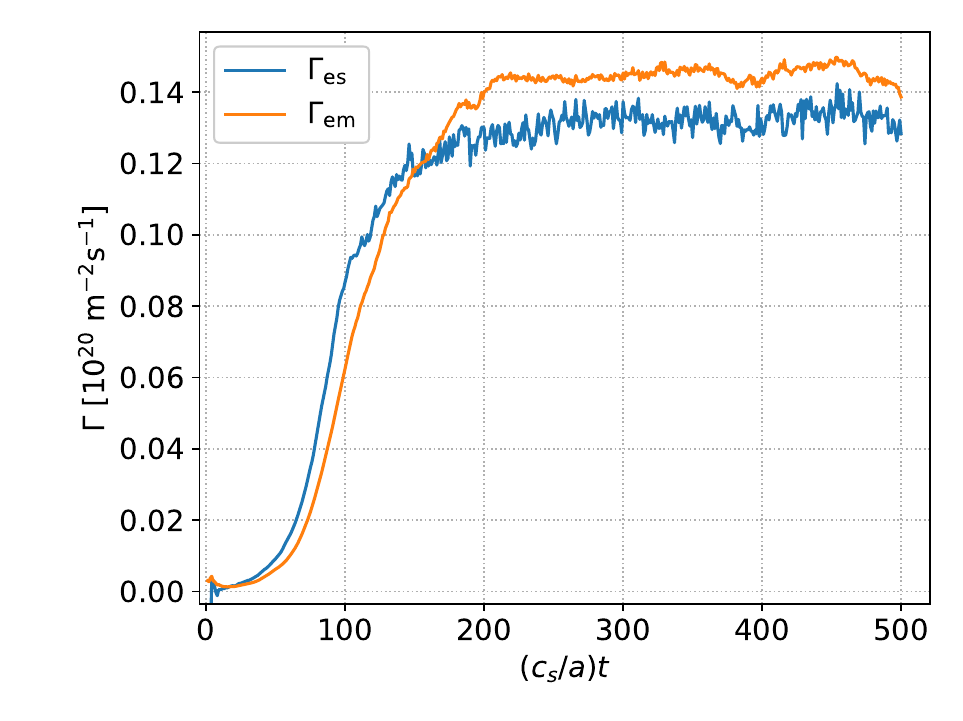}}\\
    \subfloat[]{\includegraphics[width=0.48\textwidth]{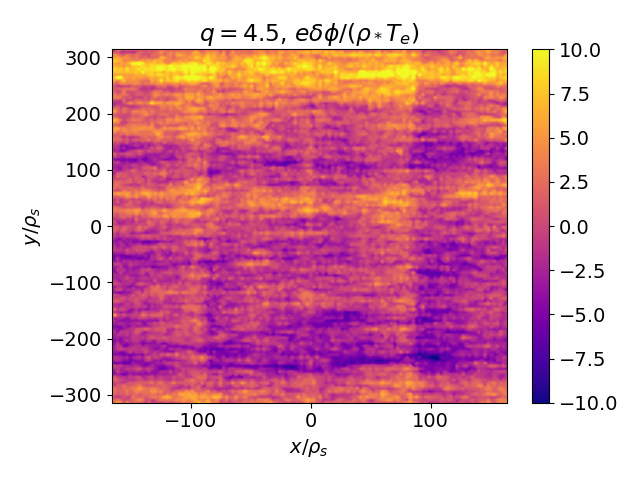}}\,
    \subfloat[]{\includegraphics[width=0.48\textwidth]{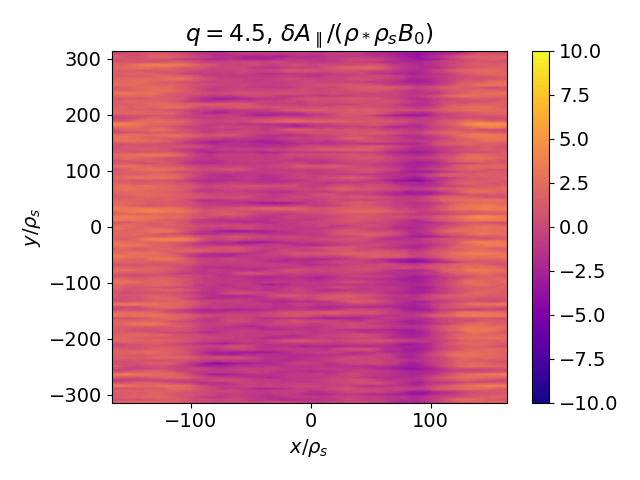}}
    \caption{Top row: time trace of the electrostatic and electromagnetic heat (a) and particle (b) fluxes from the nonlinear simulation with $q=4.5$. Bottom row: snapshot of $\delta\phi$ (c) and $\delta A_\parallel$ (d) taken in the outboard midplane at the last time step. }
    \label{fig:q45}
\end{figure}

The hypothesis that the $n=\infty$ ideal ballooning boundary can be used an indicator for the strength of the electromagnetic drive for the hybrid-KBM motivates further study of sensitivity to the safety factor.
Figure~\ref{fig:sq_scan} shows the heat and particle fluxes from a set of nonlinear simulations with different values of $q$. 
The heat and particle fluxes drop considerably when $q$ is increased and approach the target values at $q=4.5$. In all cases the heat flux is dominated by the electromagnetic electron contribution.  
We highlight that the simulations with $q=4.0$ and $q=4.5$ show a robust saturation of fluxes as well as turbulent structures that are considerably less radially extended than in the nominal case. For example, Figure~\ref{fig:q45} shows the time trace of the electrostatic and electromagnetic heat fluxes from the simulation with $q=4.5$ and snapshot contour plots of $\delta \phi$ and $\delta A_\parallel$ from the outboard mid-plane at the end of the simulation. \textcolor{black}{We also note in the simulation with $q=4.5$ that the amplitude of $\delta A_\parallel/(\rho_*\rho_sB_0)$ is smaller than the amplitude of $e\delta \phi/(\rho_*T_e)$, thus suggesting a somewhat reduced importance of electromagnetic turbulence at higher $q$.} 

\begin{figure}
    \centering
    \subfloat[]{\includegraphics[width=0.48\textwidth]{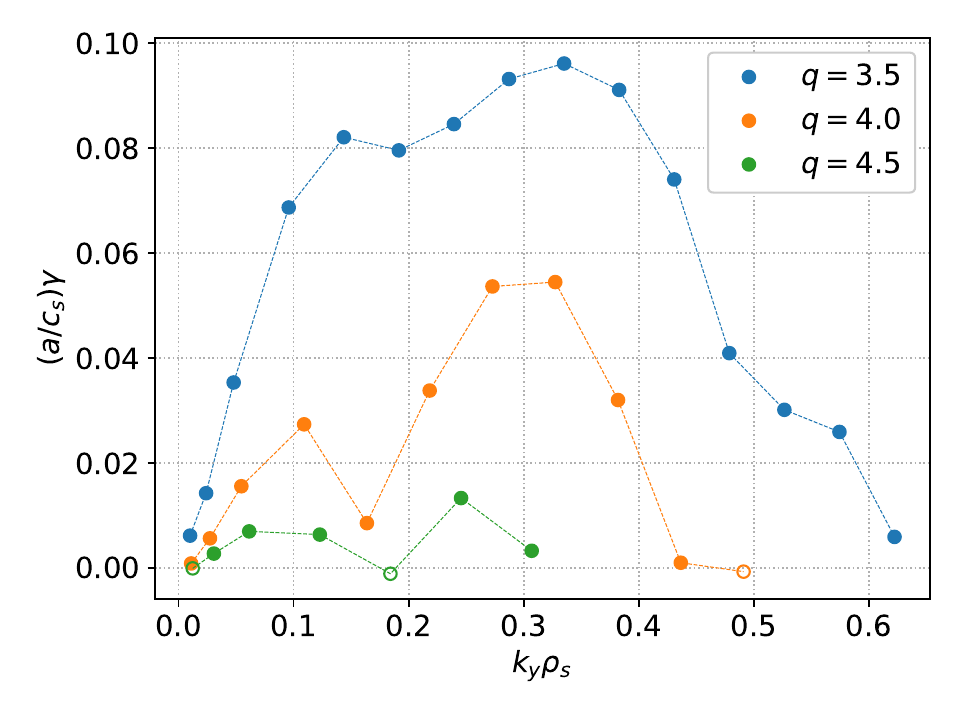}}\,
    \subfloat[]{\includegraphics[width=0.48\textwidth]{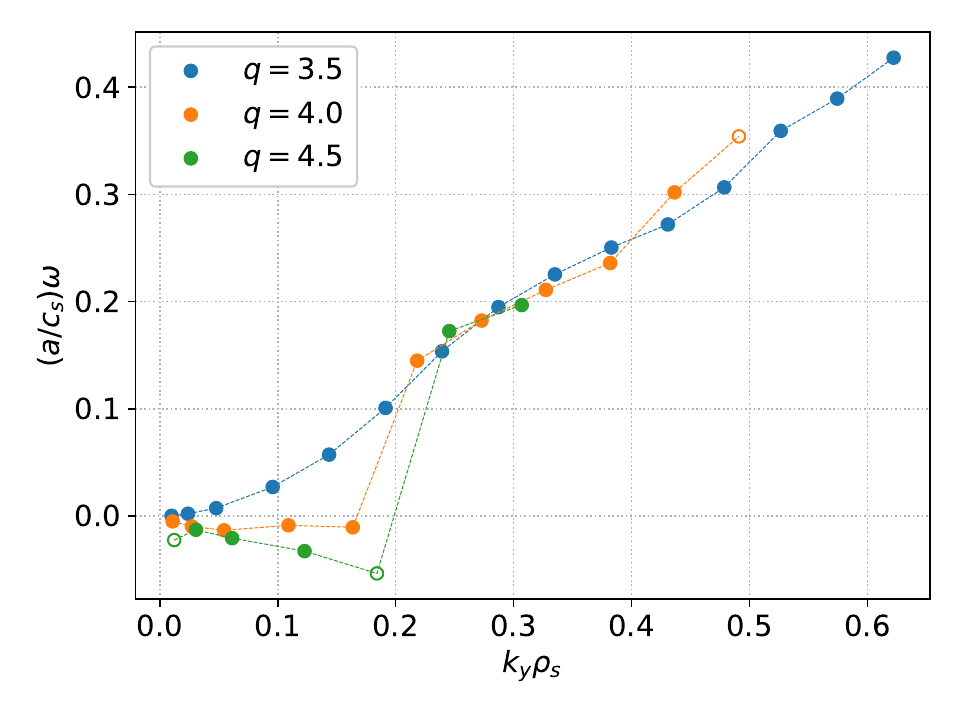}}
    \caption{Growth rate (a) and mode frequency (b) values from linear simulations at $q=3.5$ (nominal value), $q=4.0$ and $q=4.5$. Stable modes are shown with open markers.}
    \label{fig:qlinear}
\end{figure}

The nonlinear suppression of turbulent fluxes at higher $q$ illustrated of Figure~\ref{fig:sq_scan} is consistent with the strong reduction in linear growth rate values shown in Figure~\ref{fig:qlinear}.
The linear growth rate spectrum shows that a clear separation between modes at $k_y\rho_s<0.2$ and $k_y\rho_s>0.2$ appears as $q$ is increased, with a frequency sign change in the low $k_y$ region with respect to the nominal case. This may suggest that the coupling with the TEM evolves with $q$, as also mentioned in \cite{kennedy2023a}. The mode frequency at $k_y\rho_s>0.2$ is largely unaffected by the $q$ increase. 
\textcolor{black}{We highlight that, despite the strong reduction of the hybrid-KBM growth rate values at higher $q$, hybrid-KBM remains the dominant instability. In fact, the dominant mode at high $q$ has a twisting parity, the particle flux saturates at a non negligible value, and the $e\delta \phi/(\rho_*T_e)$ fluctuation amplitude is larger than the $\delta A_\parallel/(\rho_*\rho_sB_0)$ one, suggesting a the presence of an additional electrostatic turbulence drive (see Fig.~\ref{fig:q45}).}  

We note that although hybrid-KBMs are linearly very weakly unstable at $q=4.5$, they can still drive non-negligible turbulent transport nonlinearly (in the absence of equilibrium flow shear).
Similarly to the $\beta_e$ stabilisation, the beneficial effect of the safety factor may be related to an increasing distance at higher $q$ between the equilibrium point and the ideal ballooning stability boundary in the $\hat{s}-\alpha$ plane, as shown in Figure~\ref{fig:salpha}~(b). In this $q$ scan the increase in this distance at higher $q$ has a significant contribution from the movement of the boundary.

Before concluding this section, we highlight that our analysis suggests a possible strategy to mitigate transport from hybrid-KBMs when designing STEP operating points: increasing the distance between the ideal ballooning stability boundary and the equilibrium point in the $\hat{s}-\alpha$ diagram.  In all cases analysed here we find that turbulent transport is considerably reduced when the equilibrium is pushed further from the ideal ballooning limit in the direction that suppresses hybrid-KBMs in favour of other instabilities, such as TEMs or MTMs.   \textcolor{black}{In an important future work, we will perform a broader study of the  saturation mechanisms at play in hybrid-KBM turbulence, including for local equilibria further from the ideal ballooning boundary.}

\section{The role of the subdominant MTM instability}
\label{sec:subdominant MTM instability}

The subdominant MTM instability, linearly characterised in \cite{kennedy2023a}, may become important when the hybrid-KBM is suppressed and it was seen to play an interesting role as $\beta_e$ was increased in \secref{sec:beta}. It is therefore important to evaluate turbulent transport from the subdominant MTM instability, and this is computationally tractable if we can isolate unstable MTMs from the dominant hybrid-KBM.  
As shown in \cite{kennedy2023a}, the hybrid-KBM instability can be artificially suppressed {\color{black} without any impact on MTMs linearly, } by simply removing $\delta B_\parallel$ from the GK system of equations (so that only $\delta \phi$ and $\delta A_\parallel$ are evolved); this change leaves MTMs as the dominant and only instability in this reduced system. Artificial suppression of the hybrid-KBM by removing $\delta B_\parallel$ therefore provides a convenient way to study isolated MTM turbulence in the STEP operating point.  {\color{black} While it has been demonstrated that exclusion of $\delta B_{\parallel}$ has a negligible impact on the linear physics of MTMs in these STEP plasmas, the calculations of isolated MTM turbulence presented here also exclude potential nonlinear impacts associated with $\delta B_{\parallel}$ that may or may not be significant.} 

We have adopted this approach to study MTM-driven turbulence on the reference surface in STEP-EC-HD.  Whilst it is unphysical to neglect $\delta B_{\parallel}$, this study is nevertheless of considerable interest because (i) MTMs have been demonstrated to be unstable and even dominant over an extended range of binormal scales in various conceptual designs of fusion power plants based on the ST \cite{wilson2004,patel2021,dickinson2022}, (ii)  MTMs may be important  where the hybrid-KBMs is suppressed (e.g. at high $\beta^\prime$ as considered in \secref{sec:beta}), or on other radial surfaces in STEP-EC-HD, and (iii) MTMs have been demonstrated to have significant impacts on transport in other devices ~\cite{applegate2004,doerk2011,guttenfelder2011,maeyama2017,pueschel2020,giacomin2023a} and are consistent with pedestal magnetic fluctuation measurements in several experiments \cite{curie2022,kotschenreuther2019,chen2020,hatch2016,hatch2021,hassan2021,nelson2021,diallo2018,DICKINSON_PRL2012}.
Modelling low $k_y$ MTMs with conventional gyrokinetic codes is challenging due to the multiscale nature of the mode resulting in large grids and corresponding large computational cost: $\delta\phi$ is highly extended in the field-line-following coordinate $\theta,$ which implies high radial wavenumbers, while $\delta A_{\parallel}$ is localised in $\theta$ and radially extended. Due to the very high computational cost, there are only a limited number of studies of MTM-driven turbulence in the literature (see e.g. \cite{guttenfelder2011,doerk2011,giacomin2023a}). 

In our local GK simulations of MTM turbulence in STEP we find that the following numerical resolutions are sufficient: $n_\theta=32$; $n_{k_x}=512$; $n_\xi=32$; $n_v=8$; $\Delta k_y\rho_s=0.02$; $\Delta k_x = 0.04$; and 16 toroidal modes as MTMs are unstable only in a sub-region of the $k_y$ interval where the hybrid-KBM is unstable. Aside from the suppression of $\delta B_\parallel$ and the refined numerical grid, the simulation parameters are identical to those given in Table~\ref{tab:equilibrium}.
\textcolor{black}{The MTM simulation is performed with $\gamma_E = 0$.}

\begin{figure}
    \centering
    \subfloat[]{\includegraphics[width=0.48\textwidth]{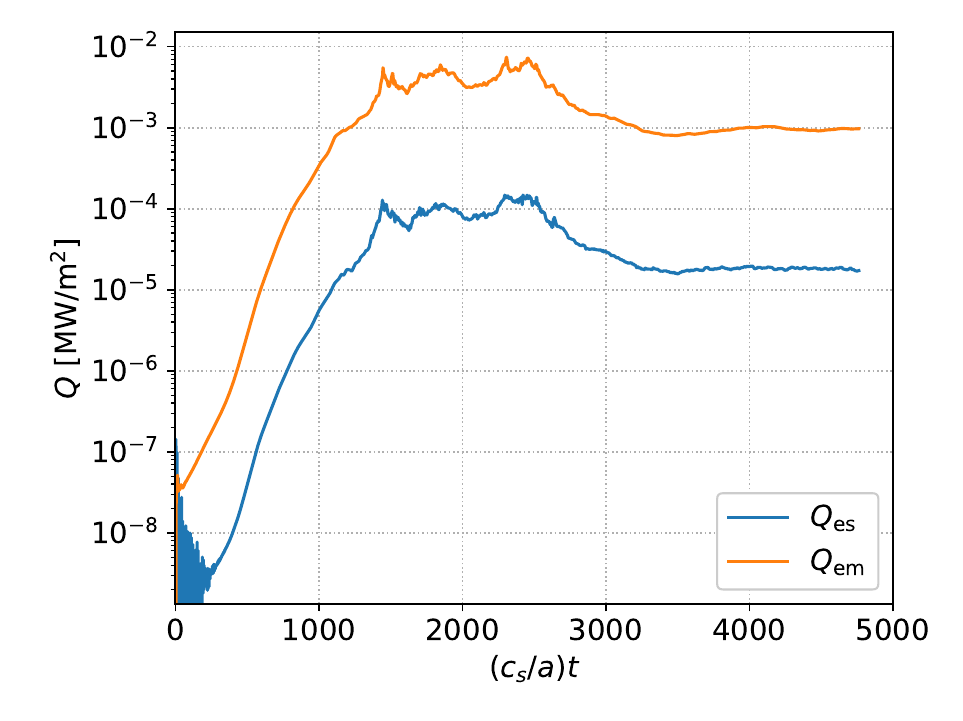}}\,
    \subfloat[]{\includegraphics[width=0.48\textwidth]{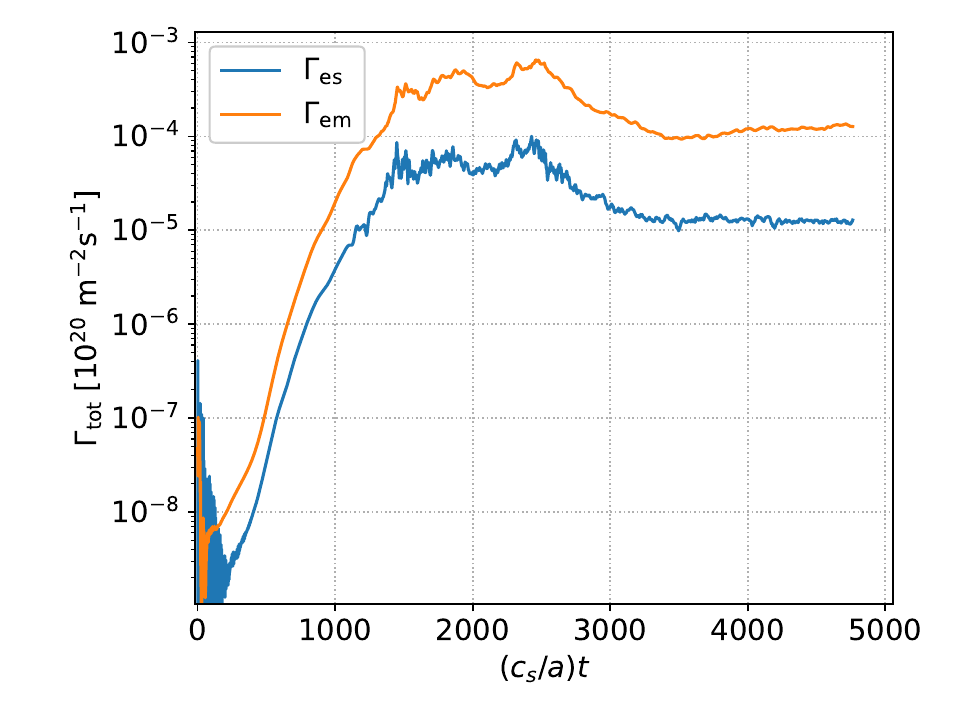}}\\
    \subfloat[]{\includegraphics[width=0.48\textwidth]{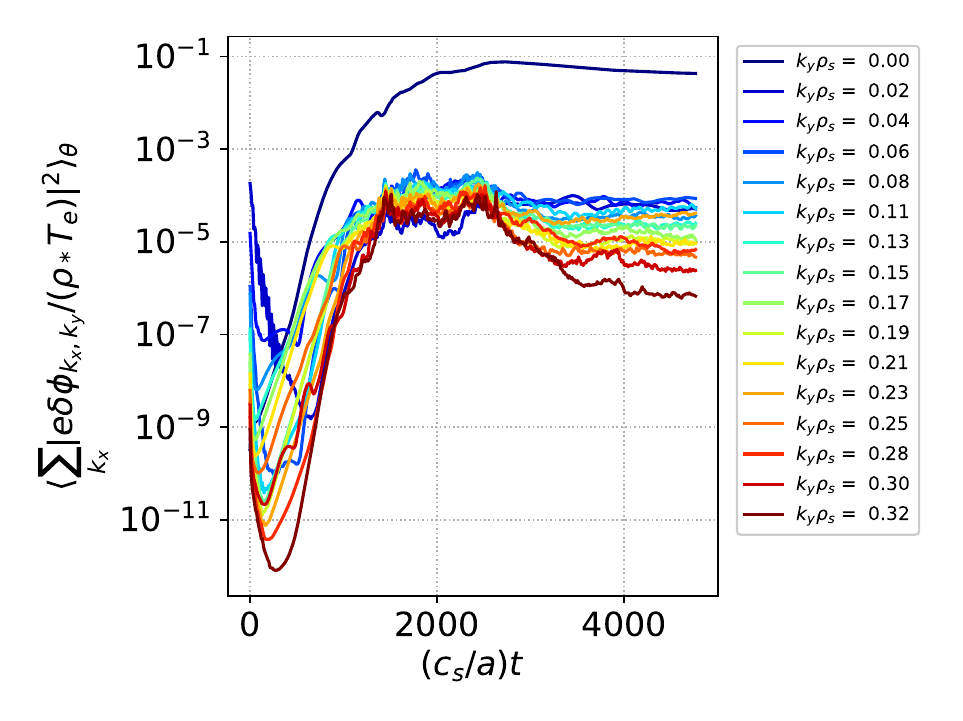}}\,
    \subfloat[]{\includegraphics[width=0.48\textwidth]{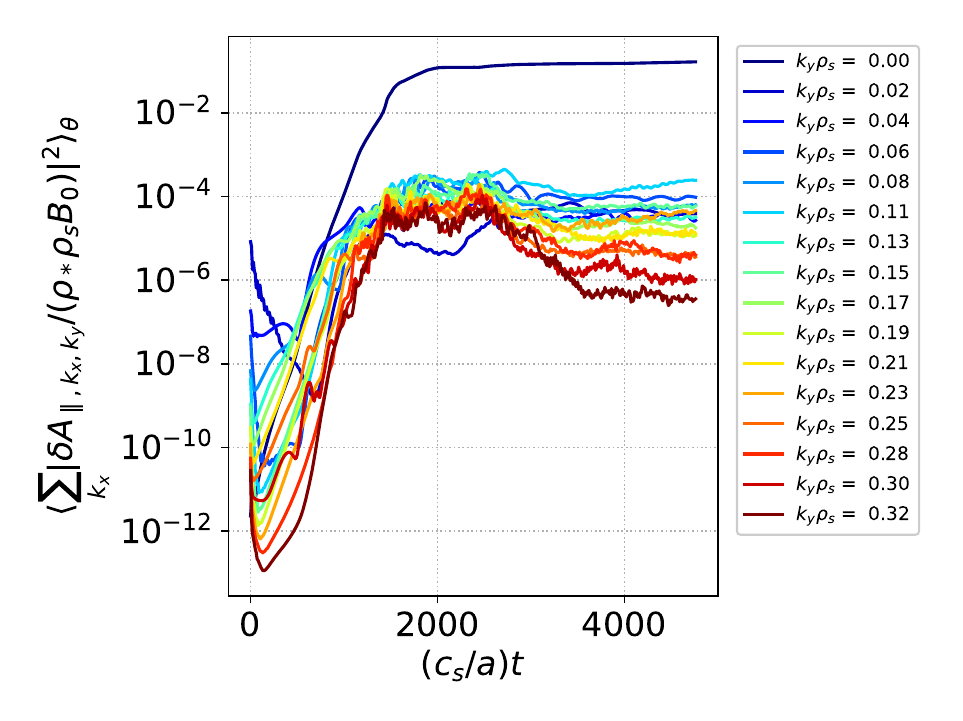}}
    \caption{Top row: time trace of the electrostatic and electromagnetic heat (a) and particle (b) fluxes from a nonlinear simulation with $\delta B_\parallel=0$ (MTM instability). Bottom row: time trace of the $k_y$ spectrum of $\delta \phi$ (c) and $\delta A_\parallel$ (d) from the same simulation.}
    \label{fig:mtm}
\end{figure}

\begin{figure}
    \centering
    \subfloat[]{\includegraphics[width=0.45\textwidth]{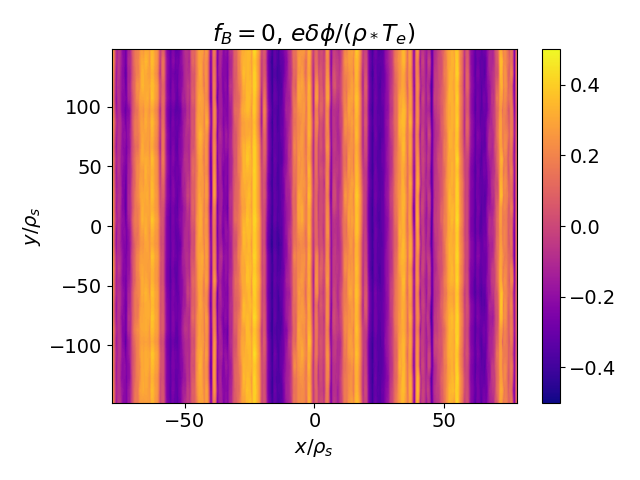}}\,
    \subfloat[]{\includegraphics[width=0.45\textwidth]{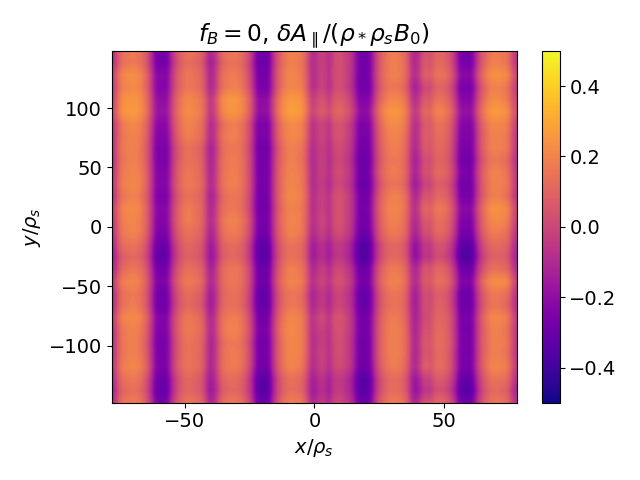}}\\
    \caption{Snapshot of $\delta \phi$ (a) and $\delta A_\parallel$ (b) taken in the outboard midplane at the last time step from the simulation without $\delta B_\parallel$.}
    \label{fig:real_mtm}
\end{figure}

\setcounter{footnote}{0}
Figure~\ref{fig:mtm} shows the time trace of the electrostatic and electromagnetic heat and particle fluxes from this nonlinear simulation, which ran for approximately fifty growth times of the most unstable $k_y$ (we note that the linear growth rate of the MTM is typically an order of magnitude smaller than that of the hybrid-KBM as shown in Figure~\ref{fig:linear}).
A robustly-steady saturated state is reached after $t\simeq 3000\,a/c_s$, where both the heat and particle fluxes are found to saturate at negligible levels\footnote{An appropriate parallel dissipation scheme was found to be essential to avoid runaway MTM fluxes \cite{dickinson2022} due to the onset of unphysical numerical instabilities with grid-scale oscillations in the parallel direction.  This finding prompted an improvement to the parallel dissipation scheme in CGYRO, implemented at commit \texttt{399deb4c}, that was used in all of the CGYRO simulations in this paper.}. 
Figure~\ref{fig:mtm} also shows the time trace of the $k_y$ spectrum of $\delta\phi$ and $\delta A_\parallel$. It can be seen that the zonal mode in both $\delta \phi$ and $\delta A_\parallel$ dominates over the non-zonal modes.
In these simulations zonal fields~\cite{pueschel2013,giacomin2023a}, and local temperature flattening~\cite{ajay2022}, are found to play a role in saturating this MTM instability at negligible heat flux values (see \ref{sec:saturation_mtm}). 
We mention that heat and particle flux predictions from the CGYRO simulation without $\delta B_\parallel$ are in good agreement with analogous simulations carried out with GENE and GS2 (see \ref{sec:comparison}).
The dominance of zonal flows and fields is clearly observed in Figure~\ref{fig:real_mtm}, which shows contour plots of $\delta \phi$ and $\delta A_\parallel$ at the outboard mid-plane taken from the last time point in the simulation. Turbulent eddies are strongly suppressed due to the presence of zonal flows and, consequently, turbulent transport is extremely modest.

It is interesting to note that although the linear growth rates of the subdominant MTM instability (see Figure~\ref{fig:linear}) are similar to the linear growth rates of the hybrid-KBM instability at higher safety factor (see Figure~\ref{fig:qlinear}), the turbulent fluxes driven by hybrid-KBMs are much larger than those driven by MTMs. This further highlights the importance of strategies to optimise the STEP equilibrium to suppress the hybrid-KBM in favour of other microinstabilities, perhaps using proximity to the ideal ballooning limit as a guide, as discussed in \secref{sec:mitigation}.

\section{Conclusions}
\label{sec:conclusions} 

This paper presents the first nonlinear gyrokinetic simulations of turbulent transport in the core plasma of the conceptual STEP reference flat-top operating point, STEP-EC-HD, based on local nonlinear simulations at the $q=3.5$  surface at $\Psi_n=0.49$.  The linear microstability analysis reported in \cite{kennedy2023a} has revealed that all unstable modes are strongly electromagnetic and at ion binormal scales (i.e. $k_y \rho_s =O(1)$), the fastest growing modes are hybrid-KBMs, and MTMs are subdominant. 

The turbulence is found to be strongly electromagnetic and dominated by hybrid-KBMs, which can drive large heat and particle fluxes that are very sensitive to equilibrium flow shear.
Simulations with $\gamma_E=0$ are characterised by radially elongated turbulent structures and very large turbulent fluxes.  
Accurately predicting turbulent transport in this regime with $\gamma_E=0$ might require global gyrokinetic simulations that can capture equilibrium profile variation. 
Although substantial progress is being made in this direction (see e.g. \cite{wilms2023}), most publicly available global codes do not, at the time of writing, retain magnetic compressional fluctuations ($\delta B_\parallel$), which are demonstrated in local gyrokinetic simulations to be essential for the hybrid-KBM to be unstable \cite{kennedy2023a}. When the hybrid-KBM is stabilised at long-wavelength  by varying local equilibrium parameters, turbulent fluxes saturate at much more reasonable levels and turbulence appears to be well-described by the local gyrokinetic model (see e.g. Figure~\ref{fig:q45}). 

Local gyrokinetic simulations that include a modest level of equilibrium flow shear saturate more robustly at considerably lower turbulent fluxes (see \secref{sec:hybrid KBM instability}). The saturated states are characterised by strong zonal flows and fields, and lower amplitude turbulent structures that are less extended radially; this turbulence appears to be better described by local gyrokinetics. Even in devices such as STEP, where rotation is expected to be modest due to the lack of external momentum injection, diamagnetic levels of flow shear can play an important role in saturating hybrid-KBMs. 

Robustly saturated simulations at lower fluxes can also be obtained at $\gamma_E=0$ when the local equilibrium parameters act to suppress the linear drive for hybrid-KBMs. A study of the sensitivity of turbulent fluxes to the pressure gradient in \secref{sec: sensitivity to local parameters} reveals a crucial competition between the linear drive of hybrid-KBMs and the stabilising impact of $\beta^\prime$ when $\beta^{\prime}$ is varied consistently.  Strikingly, in the region of this local reference equilibrium, $\beta^{\prime}$ stabilisation overpowers the drive, and turbulent fluxes reduce or increase substantially when the pressure gradient is increased or reduced, respectively. This suggests that a favourable confinement regime exists at high $\beta^{\prime}$, but accessing this regime would require crossing, or better evading, higher transport states at lower $\beta^{\prime}$.
Further sensitivity studies reveal that either increasing $\beta_e$ consistently with $\beta^{\prime}$, or increasing $q$, leads to a substantial reduction in the hybrid-KBM turbulent fluxes, and expose potential transport mitigation mechanisms to supplement flow shear. These reductions in transport are found to correlate with distance between the equilibrium point and the ideal ballooning boundary in the $\hat{s}-\alpha$ diagram: at larger distances the transport fluxes and hybrid-KBM linear growth rates are reduced.  This offers a simple approach to guide the optimisation of STEP scenarios, though it cannot capture important details of the turbulent transport\footnote{The low and high $\beta_e$ points in Figure~\ref{fig:salpha}~(a) which are at similar distances from the ideal ballooning boundary, but are in very different turbulent transport regimes as the low $\beta_e$ case is dominated by electrostatic turbulence while the high $\beta_e$ case is dominated by MTM turbulence}. Interestingly, we also note that heat and particle fluxes become more compatible with the available STEP sources when the local equilibrium is pushed to a regime where hybrid-KBMs are suppressed in favour of MTMs.  Exploiting insights like those developed here will be essential for establishing routes to more optimal transport regimes.

Isolated turbulence from the linearly sub-dominant MTMs is also modelled in~\secref{sec:subdominant MTM instability}, by artificially suppressing  hybrid-KBMs through the neglect of  $\delta B_{\parallel}$  fluctuations. The MTM-driven heat flux saturates at negligible transport levels on the chosen surface in STEP-EC-HD, suggesting we should seek regimes where the hybrid-KBM is more stable, even if this comes at the expense of increasing the drive for MTMs.  

In conclusion, first nonlinear local gyrokinetic simulations for a mid-radius surface in STEP-EC-HD predict that the plasma sits in a novel and poorly understood turbulent transport regime dominated by strongly electromagnetic hybrid-KBM turbulence.  In the absence of equilibrium flow shear, simulations predict transport fluxes that \textcolor{black}{are well above values compatible with STEP sources}. Nevertheless, turbulent fluxes can be significantly reduced by including perpendicular flow shear (even at the diamagnetic level), and by other changes to the local equilibrium that increase distance from the $n=\infty$ ideal ballooning boundary  (e.g. higher $\beta^{\prime}$ and higher $q$).  Local gyrokinetics appears to be more suitable for describing the turbulence in these mitigated states.  These first calculations will play an important role in informing the future optimisation of STEP.

\textcolor{black}{The analysis performed here  exposes a clear need to account for hybrid-KBM driven turbulent transport in future efforts to optimise STEP plasmas, which will likely require several avenues to be pursued in parallel. Firstly, there is a clear need to improve the fidelity of hybrid-KBM turbulence simulations through the inclusion of physics likely to impact on the turbulent fluxes that has been neglected here, such as global physics, impurities, and fast particles. Secondly, it is urgent to hone the best available model(s) of hybrid-KBM turbulence and exploit them to predict the transport steady state; e.g. through computationally challenging coupled transport-turbulence simulations using local GK to model the hybrid-KBM turbulence or using higher fidelity reduced transport model able to capture hybrid-KBMs.
A third priority is to improve our understanding of the basic physics associated with the onset of large fluxes as well as of the hybrid-KBM saturation mechanisms.
Finally, it will be important to validate calculations of hybrid-KBM turbulence in detail against data from experiments in STEP-relevant regimes, i.e. in high $\beta$, low collisionality, low torque plasmas with substantial electron heating. We note that MAST-U will install additional NBI power and an Electron Bernstein Wave heating and current drive system~\cite{webster2023}, which should produce ST plasmas in regimes more relevant to STEP than it has been achieved previously.}
While these activities are important for advancing the ST route to a fusion power plant, they may also be useful for other devices seeking to operate in similar regimes.

\section*{Acknowledgements}

The authors would like to thank E. Belli and J. Candy for their very helpful support with CGYRO simulations as well as T. G{\"o}rler and D. Told for their help with GENE. The authors would also like to thank B. Chapman-Oplopoiou, M. Hardman, D. Hatch, F. Sheffield Heit, and P. Ivanov, and  for helpful discussions and suggestions at various stages of this project. D. Kennedy is grateful to The Institute for Fusion Studies (IFS), Austin TX, for its splendid hospitality during stimulating and productive visits during the course of this work. This work has been funded by the Engineering and Physical Sciences Research Council (grant numbers EP/R034737/1 and EP/W006839/1). Simulations have been performed on the ARCHER2 UK National Supercomputing Service under the project e607 and on the Marconi National Supercomputing Consortium CINECA (Italy) under the projects STETGMTM and QLTURB. Part of this work was performed using resources provided by the Cambridge Service for Data Driven Discovery (CSD3) operated by the University of Cambridge Research Computing Service (\url{www.csd3.cam.ac.uk}), provided by Dell EMC and Intel using Tier-2 funding from the Engineering and Physical Sciences Research Council (capital grant EP/T022159/1), and DiRAC funding from the Science and Technology Facilities Council (\url{www.dirac.ac.uk}).

\appendix
\section{Impact of numerical resolution}
\label{sec:numerical_convergence}

The analysis in \secref{sec:hybrid KBM instability} shows that, in absence of equilibrium flow shear, turbulent fluxes reach large values with no signature of having attained a robustly-steady saturated state and with long-lived turbulent structures that are radially extended. In this appendix, we show that this state of very large fluxes is \emph{not} a numerical feature due to numerical resolution. A convergence study with respect to the radial box size is also presented for simulations with shearing rate of $\gamma_E=0.1\,c_s/a$. 

\subsection{Slightly reduced base grid in $k_y$ for resolution studies}
\begin{figure}
    \centering
    \subfloat[]{\includegraphics[height=0.23\textheight]{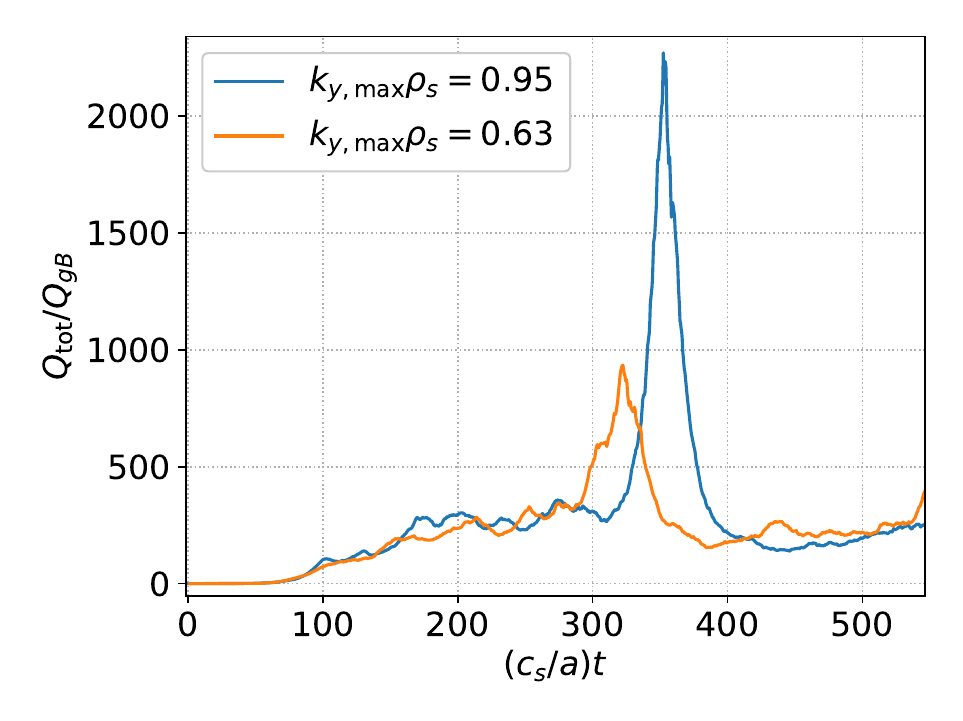}}\,
    \subfloat[]{\includegraphics[height=0.23\textheight]{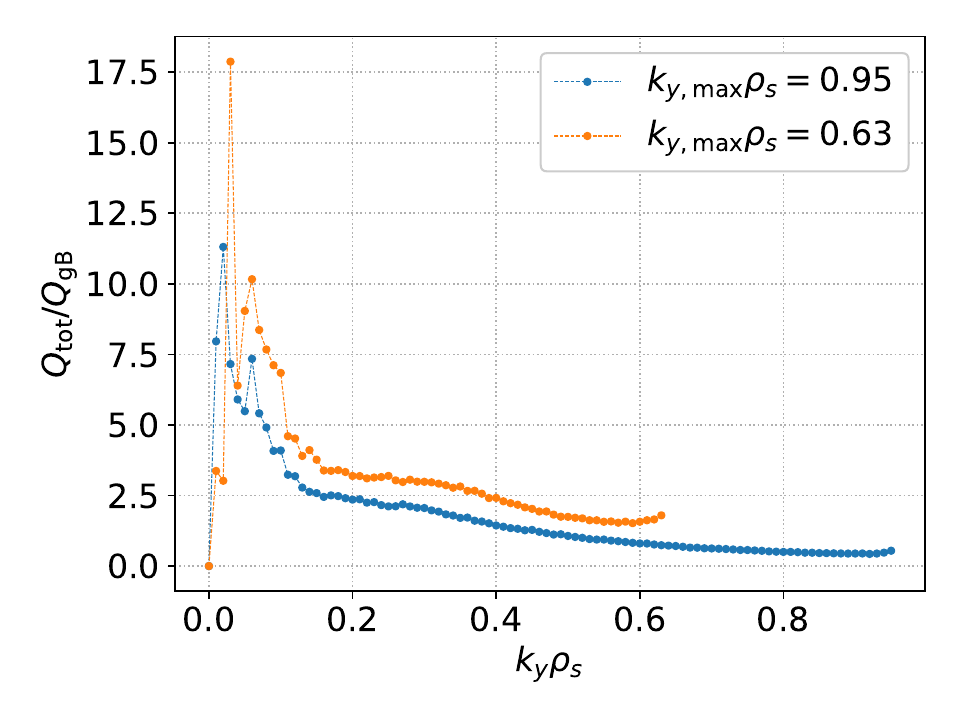}}
    \caption{(a) Time trace of the total heat flux from nonlinear simulations with $k_{y,\mathrm{max}}\rho_s=0.95$ and $k_{y,\mathrm{max}}\rho_s=0.63$. (b) $k_y$ spectrum of the total heat flux from the same simulations averaged over the time interval $t\in [400, 500]\,a/c_s$.}
    \label{fig:nky}
\end{figure}

In order to reduce the computational cost of these resolution test simulations, the value of  $k_{y,\mathrm{max}}$ is decreased from $k_{y,\mathrm{max}}\rho_s = 0.95$ used in the main text (e.g. see Figure~\ref{fig:timetrace}) to $k_{y,\mathrm{max}}\rho_s=0.63$ for the simulations in this appendix. This value of $k_{y,\mathrm{max}}$ is the largest $k_y$ value at which hybrid-KBMs are unstable (see Figure~\ref{fig:linear}). In addition, as shown in Figure~\ref{fig:timetrace}, modes with $k_y\rho_s > 0.6$ give negligible contributions to the total heat flux.
Figure~\ref{fig:nky} compares the time trace of the total heat flux of the nominal simulation (see \secref{sec:hybrid KBM instability}) and of a simulation with $k_{y,\mathrm{max}}\rho_s=0.63$. The time traces shows good quantitative agreement before $t\simeq 300\,a/c_s$, both reaching similarly large heat flux values in this timeframe. 
Around $t\simeq 300\,a/c_s$, both simulations show heat flux bursts that are qualitatively similar but differ in magnitude. After the heat flux burst, from $t\gtrsim 400\,a/c_s$ there is better agreement and the heat flux values remain large at $200-300\,Q_{gB}$.
Therefore, apart from the heat flux burst, the two simulations agree quite well. 
Figure~\ref{fig:nky} compares also the $k_y$ spectra of the total heat flux averaged over the time interval $t\in [400, 500]\,a/c_s$, and both spectra are similar at low $k_y$. Some quantitative difference is observed at $k_y\rho_s \gtrsim 0.2$, where the heat flux values are slightly larger in the case of $k_{y,\mathrm{max}}\rho_s=0.63$, while the total contribution from $k_y\rho_s \gtrsim 0.2$ is very similar in the two cases.
In order to reduce the computational cost of these test simulations, we use $k_{y,\mathrm{max}}\rho_s=0.63$ in this appendix, keeping the same value of $k_{y,\mathrm{min}}$ to properly resolve the (most critical) low $k_y$ modes. 

\subsection{Radial box size}
In flux-tube simulations that make use of ``twist-and-shift'' parallel boundary conditions~\cite{beer1995}, the radial box size, $L_x,$ is determined from $k_{y, \mathrm{min}}$ and $\hat{s}$ by the equation $L_x=j/(\hat{s}k_{y, \mathrm{min}})$, where $j$ is an integer.
Once $k_{y, \mathrm{min}}$ has been chosen, the value of $L_x$ is solely determined by $j$.
Here, we consider three simulations with different values of $j$ and $n_{k_x}$: (i) $j=2$ and $n_{k_x}=64$, (ii) $j=4$ and $n_{k_x}=64$, and (iii) $j=8$ and $n_{k_x}=128$. The numerical resolution of the other parameters is the same as in the nominal case, except for the reduced value of $k_{y,\mathrm{max}}$. 

The value of $L_x$ is doubled from (i) to (ii) while keeping the same number of radial grid points, i.e. both $\Delta k_x$ and $k_{x, \mathrm{max}}$ are reduced by a factor of two. The value of $L_x$ is doubled again from (ii) to (iii), while doubling also the value of $n_{k_x}$, i.e.  $\Delta k_x$ is reduced by a factor of two relative to (ii) and $k_{x, \mathrm{max}}$ is kept constant. 
Figure~\ref{fig:resolution} shows the time trace of the total heat and particle fluxes from these three simulations. In all cases, turbulent fluxes reach very large values, with significant oscillations appearing at all the considered resolutions: there is no improvement when a larger radial domain is considered and turbulent eddies that are similar to the nominal case shown in Figure~\ref{fig:real_exb0} and scale with the radial box size. 

\subsection{Initial amplitudes, \texorpdfstring{$n_{\theta}$}{TEXT} and \texorpdfstring{$n_{\xi}$}{TEXT}, and different local GK codes}

\begin{figure}
    \centering
    \subfloat[]{\includegraphics[height=0.23\textheight]{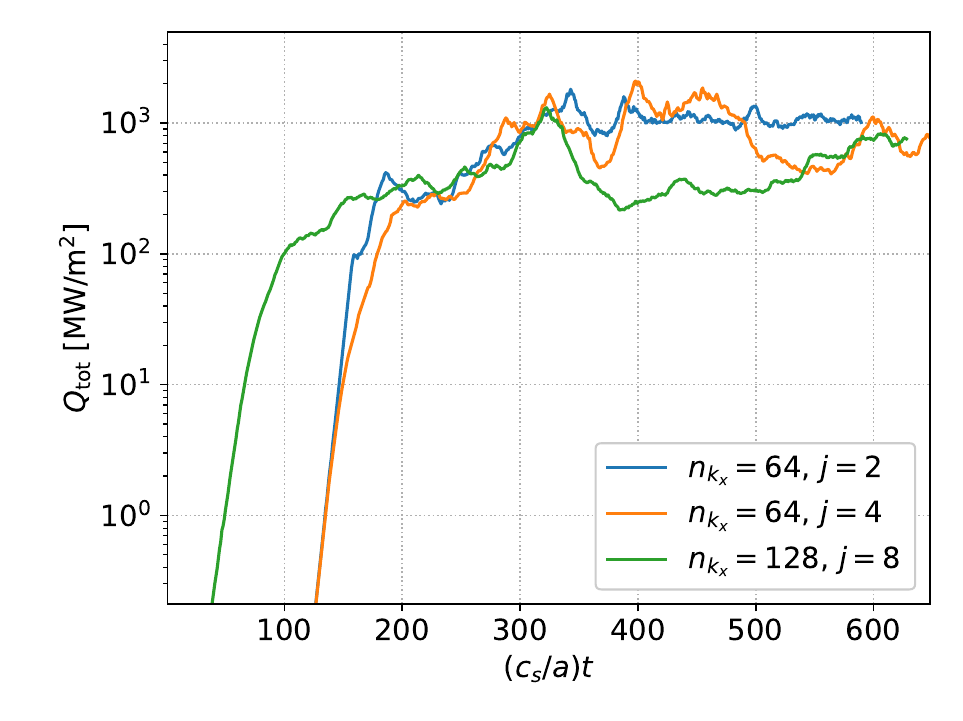}}\,
    \subfloat[]{\includegraphics[height=0.23\textheight]{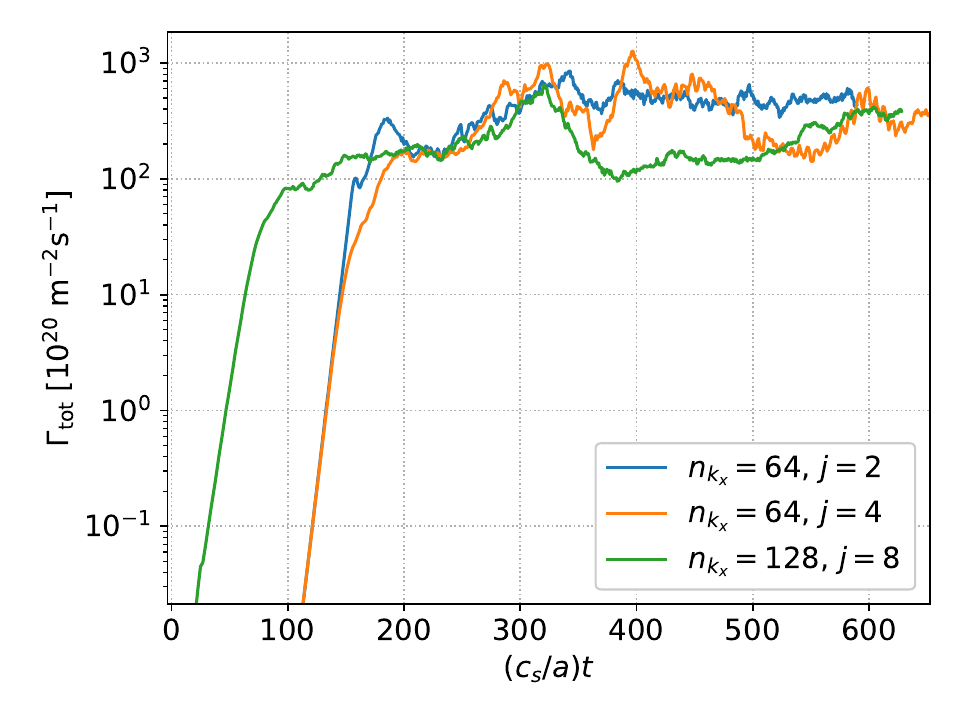}}
    \caption{Time trace of the total heat (a) and particle (b) flux from nonlinear simulations with different radial domain size (controlled via $j$) and radial resolution (controlled via $n_{k_x}$). The initial condition amplitude of the simulations with $n_{k_x}=64$ is smaller than the one of the simulation with $n_{k_x}=128$.  }
    \label{fig:resolution}
\end{figure}

Different initial amplitudes are applied in the simulations shown in Figure~\ref{fig:resolution}, but no impact of the initial condition on the saturated turbulence is observed. 

Nonlinear simulations with higher $n_\theta$ and $n_\xi$ have been also performed, and they all show the same trend and reach heat flux values of the order of $10^{3}$~MW/m$^2$. 

A similar state with very large turbulent fluxes is observed in analogous GENE simulations (see \ref{sec:comparison} for code comparison).

Indeed, all the numerical scans performed at nominal STEP flat-top parameters indicate that the large fluxes and lack of clean saturation for $\gamma_E=0$ are not artefacts related to poor resolution or other numerical issues.

\subsection{Box size for simulations with flow shear} 
\begin{figure}
    \centering
    \subfloat[]{\includegraphics[height=0.23\textheight]{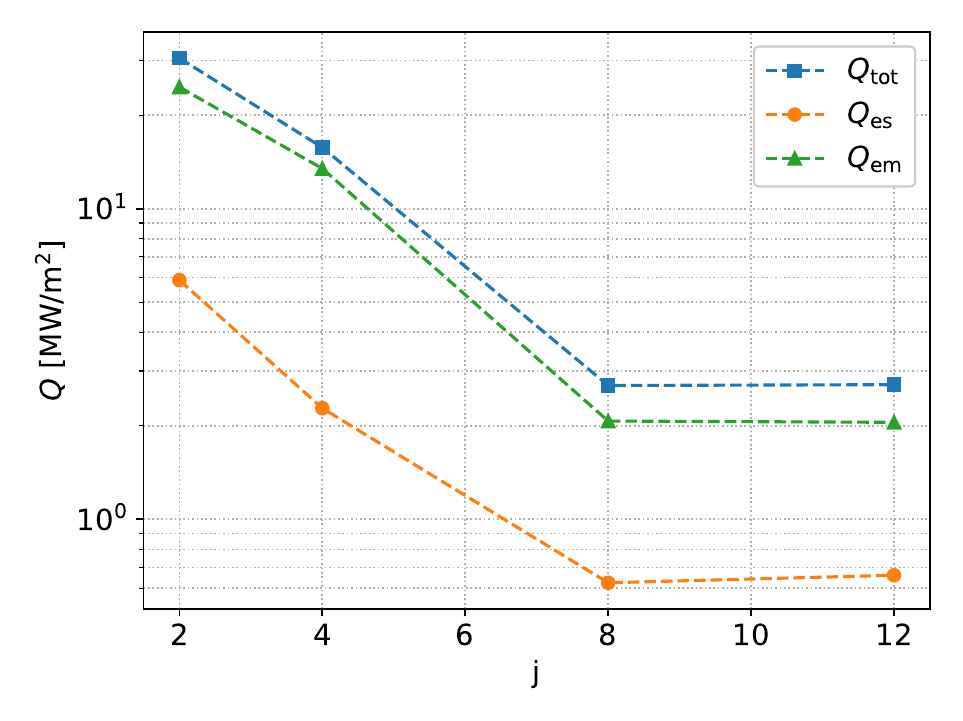}}\,
    \subfloat[]{\includegraphics[height=0.23\textheight]{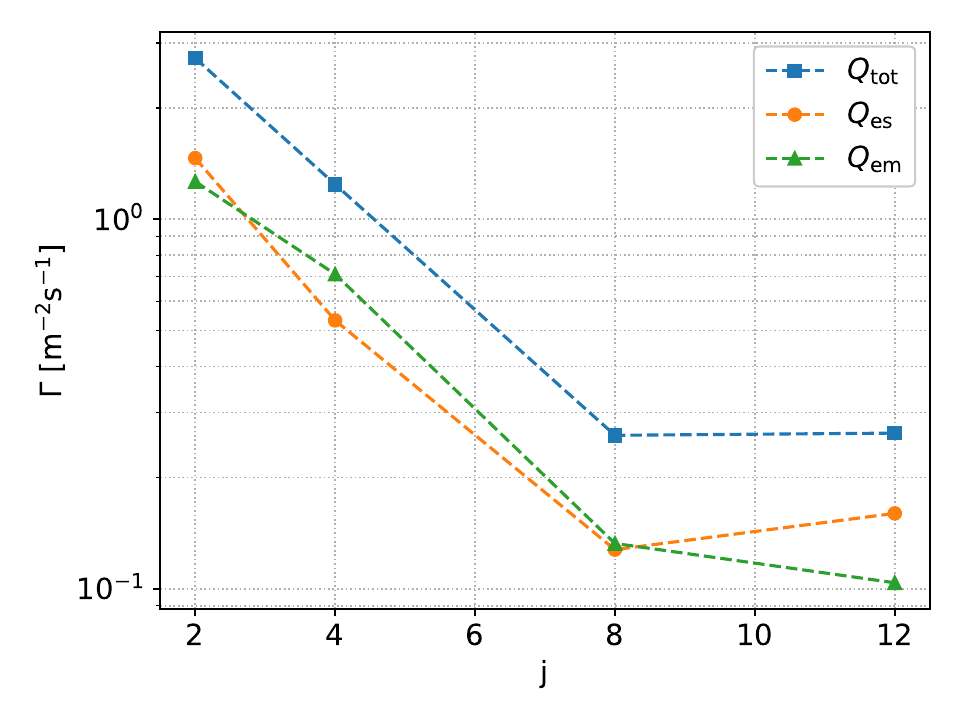}}
    \caption{Saturated value of the electromagnetic and electrostatic heat (a) and particle (b) fluxes from a set of simulations with $\gamma_E=0.1\,c_s/a$ and various radial box size (controlled via the parameter $j$). The number of radial and binormal grid points is kept fixed at $n_{k_x}=128$ and $n_{k_y}=64$.}
    \label{fig:res_exb}
\end{figure}

Considering now simulations with equilibrium flow shear, Figure~\ref{fig:res_exb} shows the saturated value of heat and particle fluxes from a set of nonlinear simulations with $\gamma_E=0.1\,c_s/a$ and various radial box size (controlled via the parameter $j$). 
The number of radial grid points is kept constant while changing the radial box size. Similarly to the numerical scan presented above, the value of $n_{k_y}$ is reduced from 96, considered in  \secref{sec:hybrid KBM instability}, to 64, considered in this appendix.
The simulations are performed with $n_{k_x}=128$ and $n_{k_y}=64$.
Convergence of heat and particle fluxes is achieved at $j\geq 8$, which is the value considered for the simulations discussed in \secref{sec:hybrid KBM instability}.

\section{Code comparison}
\label{sec:comparison}

In this appendix we briefly compare the turbulent transport predictions for STEP from three different well established local gyrokinetic codes: CGYRO \cite{candy2016} (commit \texttt{399deb4c}), GENE \cite{gene} (commit \texttt{de99981}), and GS2 \cite{gs2} (commit \texttt{675f0870}). We remark that obtaining high-calibre quantitative agreement is incredibly difficult in these highly-shaped, high-$\beta$ plasmas.  Our main objective here is simply to check that all codes describe, with a reasonable level of agreement, the same key qualitative properties of the turbulence. 
We note that an acceptable agreement on the linear growth rate values is observed among the three codes (see Figure 19 and Figure 20 of \cite{kennedy2023a}).

\subsection{Comparison of the hybrid-KBM transport}

We compare CGYRO and GENE results from nonlinear simulations with three different values of equilibrium flow shear, 
$\gamma_E \in \{0, 0.05, 0.1, 0.2\} \,c_s/a$.
The aim of this comparison is to assess whether both codes \textcolor{black}{reach a robust saturated state at $\gamma_E=0$} and find the same sensitivity of the turbulent hybrid-KBM heat flux to $\gamma_E$. 

It is important to remark that CGYRO and GENE use a different implementation of the equilibrium flow shear. 
CGYRO favours a spectral approach to impose a radially periodic equilibrium flow~\cite{candy2018}, whereas GENE implements a radially constant $\mathbf{E}\times\mathbf{B}$ shearing rate by shifting the $k_x$ grid in time~\cite{hammett2006,mcmillan2019}.  GENE uses the wavenumber remapping method and includes continuous shearing in the nonlinear term detailed in \cite{mcmillan2019} that had been neglected in earlier implementations. The GENE simulations in this appendix are performed using the Sugama collision operator (also used in CGYRO) and the following grid resolutions: $n_\theta = 32$, $n_{k_x} = 128$, $k_{y,\mathrm{min}}\rho_s = 0.01$, $n_{k_y}=64$, $n_v=32$ (number of grid points in the $v_\parallel$ direction), and $n_\mu=16$ (number of grid points in the $\mu=v_{\perp}^2/(2B)$ direction). This resolution is identical to the one used in CGYRO simulations, except for the reduced value of $n_{k_y}$, where $n_{k_y} = 96$ is used in the main text (see section.~\secref{sec:resolution}), while  $n_{k_y} = 64$ is used in the GENE simulations in this appendix.  

\begin{figure}
    \centering
    \subfloat[]{\includegraphics[height=0.19\textheight]{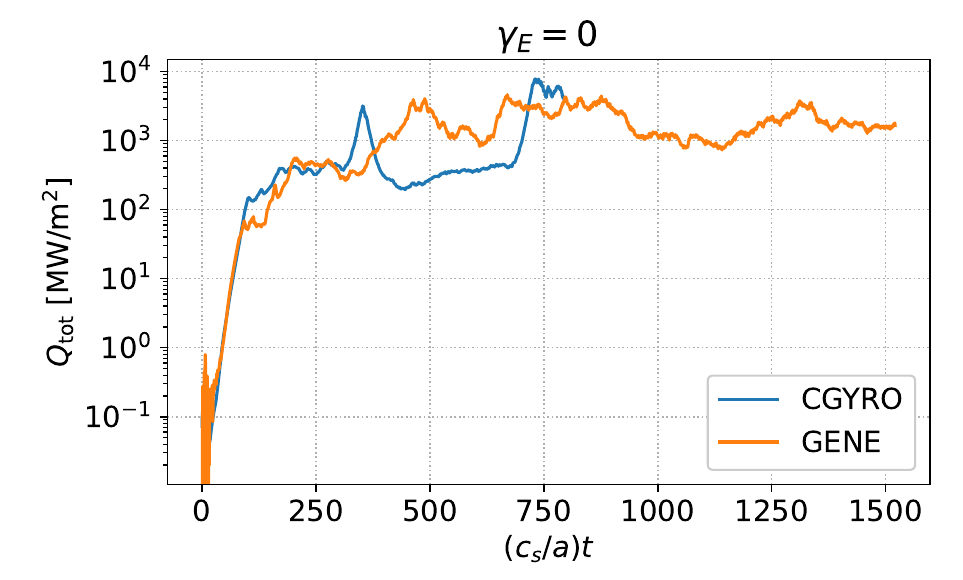}}\,
    \subfloat[]{\includegraphics[height=0.19\textheight]{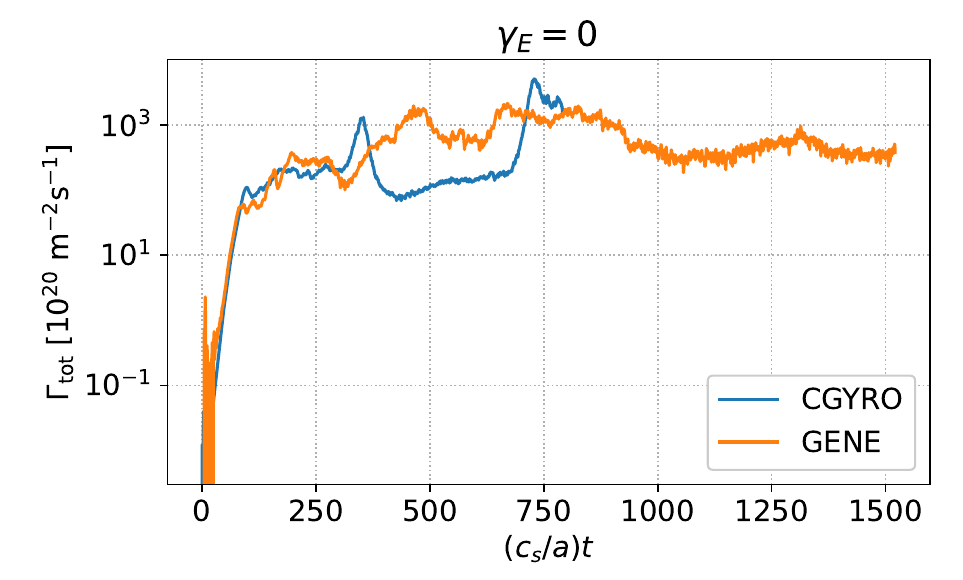}}
    \caption{Time trace of the total heat (a) and particle (b) fluxes from the CGYRO and GENE simulation with $\gamma_E=0$.}
    \label{fig:comp_gamma0}
\end{figure}

\textcolor{black}{Figure~\ref{fig:comp_gamma0} shows the time trace of the total heat and particle flux from the CGYRO and GENE simulation with $\gamma_E=0$. Despite the large turbulent flux fluctuations characterising these time traces, both simulations reach a state of stationary turbulent fluxes at similar values, thus providing further evidence that all the STEP simulations presented in this work do eventually saturate if they are run for sufficiently long time. On the other hand, we highlight that the saturated value found in CGYRO and GENE in absence of equilibrium flow shear is three orders of magnitude larger than a value that would be compatible with the STEP sources computed from JETTO.} 

\begin{figure}
    \centering
    \subfloat[]{\includegraphics[height=0.23\textheight]{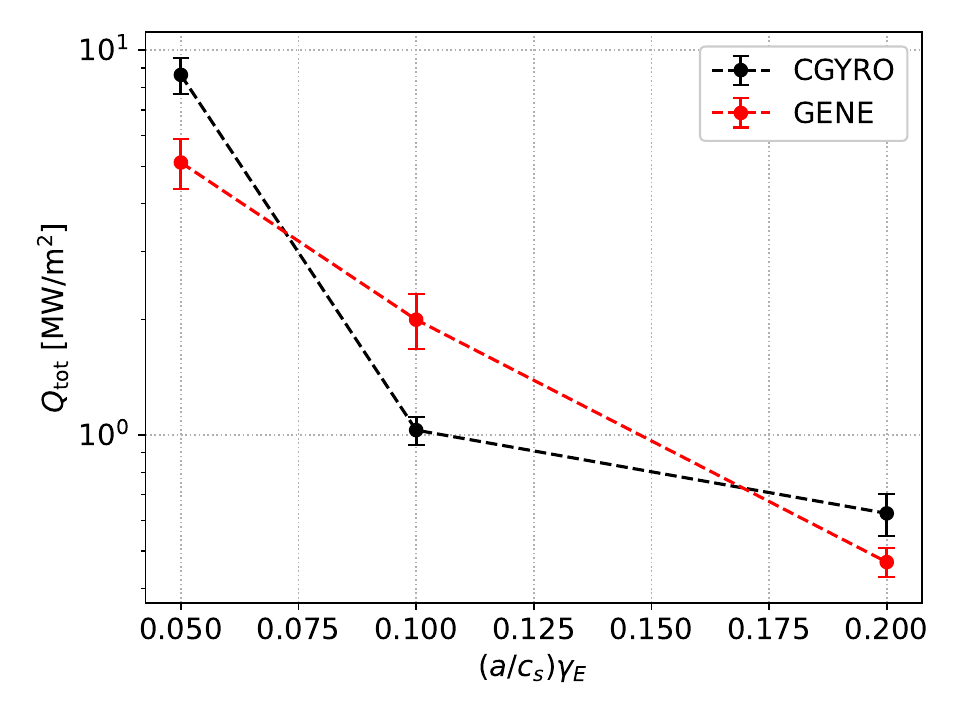}}\,
    \subfloat[]{\includegraphics[height=0.23\textheight]{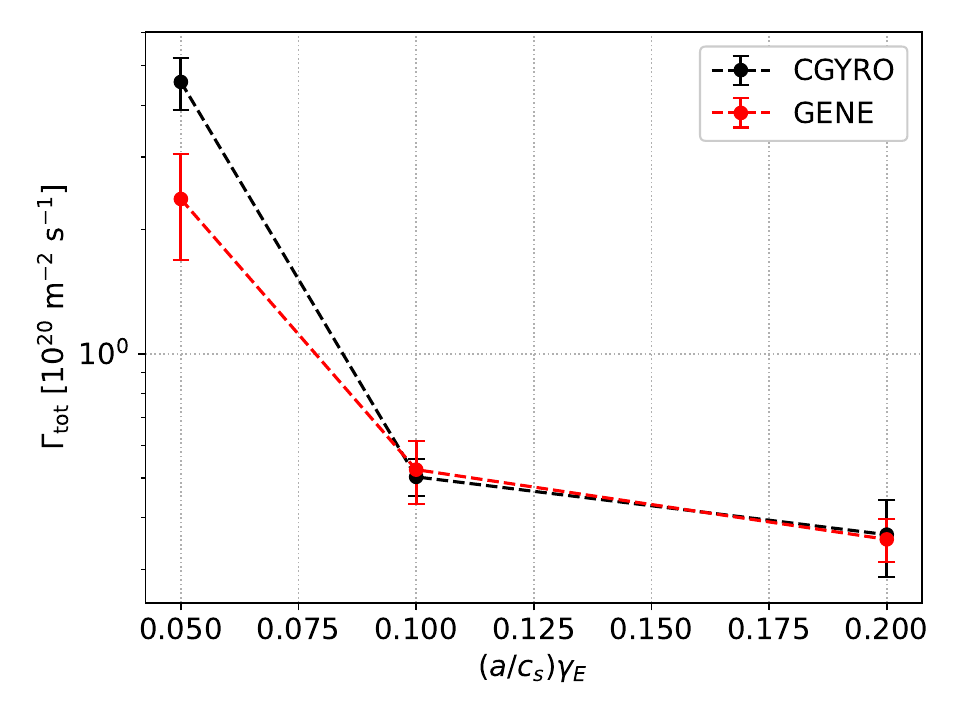}}
    \caption{Comparison between CGYRO and GENE total heat (a) and particle (b) fluxes as a function of equilibrium flow shear.}
    \label{fig:comp_cgyro_gene}
\end{figure}

Focusing now on the $\gamma_E$ sensitivity, Figure~\ref{fig:comp_cgyro_gene} shows the saturated heat and particle fluxes as functions of $\gamma_E$ from CGYRO 
and GENE. Both codes find a strong reduction in the total heat flux with increasing $\gamma_E$. There is a very good agreement between CGYRO and GENE particle flux values, and slightly poorer agreement in the heat flux, where the values agree to within a factor two.  This discrepancy could well be attributed to the very different equilibrium flow shear implementations in the codes.

\subsection{Subdominant MTM transport from CGYRO, GENE and GS2}

We compare here the heat and particle fluxes predicted by CGYRO, GENE and GS2 nonlinear simulations without $\delta B_\parallel$ to isolate turbulent transport from subdominant MTMs. 
Both GENE and GS2 simulations are carried out using numerical resolutions close to those used in the corresponding CGYRO simulation, $n_\theta = 32$, $n_{k_x}=512$, $n_{k_y}=16$, $\Delta k_y\rho_s=0.02$, $\Delta k_x = 0.04$,  $n_v = 32$ and $n_\mu = 16$ (in GENE), and $n_\epsilon = 8$ and  $n_\xi=24$ (in GS2). 
The Sugama collision operator is used in the GENE simulation, while the linearised Fokker-Planck collision model of \cite{barnes2009} is used in the GS2 simulation.

\begin{figure}
    \centering
    \subfloat[]{\includegraphics[height=0.23\textheight]{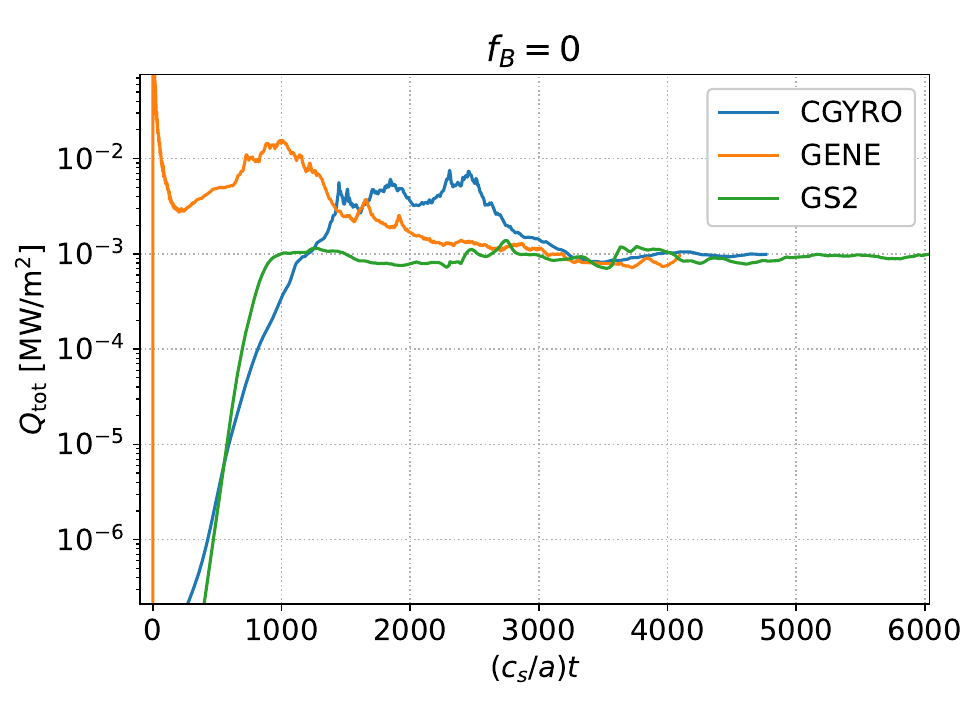}}\,
    \subfloat[]{\includegraphics[height=0.23\textheight]{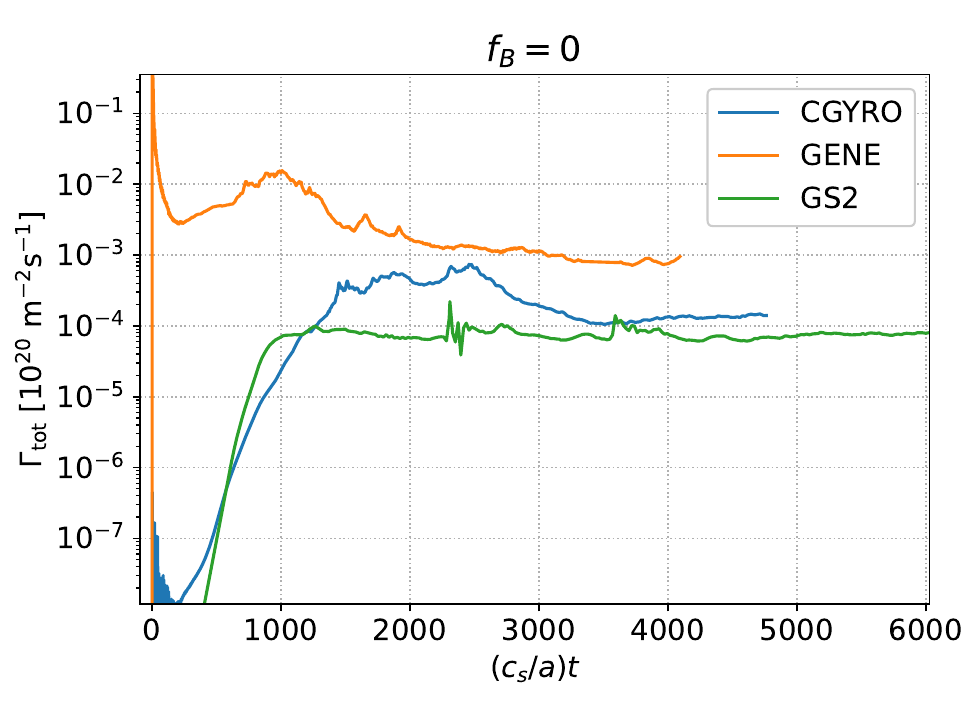}}
    \caption{Time trace of the total heat (a) and particle (b) flux from CGYRO, GENE and GS2 nonlinear simulations without $\delta B_\parallel$ (subdominant MTM instability).}
    \label{fig:comp_mtm}
\end{figure}

Figure~\ref{fig:comp_mtm} shows the time trace of the heat and particle fluxes from the simulations. The saturated level of the total heat flux from the three codes agrees very well, although the initial phase is different because of different initial conditions. The saturated heat flux level is approximately $10^{-3}$~MW/m$^{2}$ in all the three simulations.  The particle flux is negligible.  All codes agree on the most important point that the MTM transport fluxes are essentially negligible. A higher level of quantitative agreement between codes is very difficult to achieve since this case is very close to marginal stability, with very low turbulent fluxes. 

\section{Saturation mechanisms of the MTM instability}
\label{sec:saturation_mtm}
\begin{figure}
    \centering
    \includegraphics[scale=0.6]{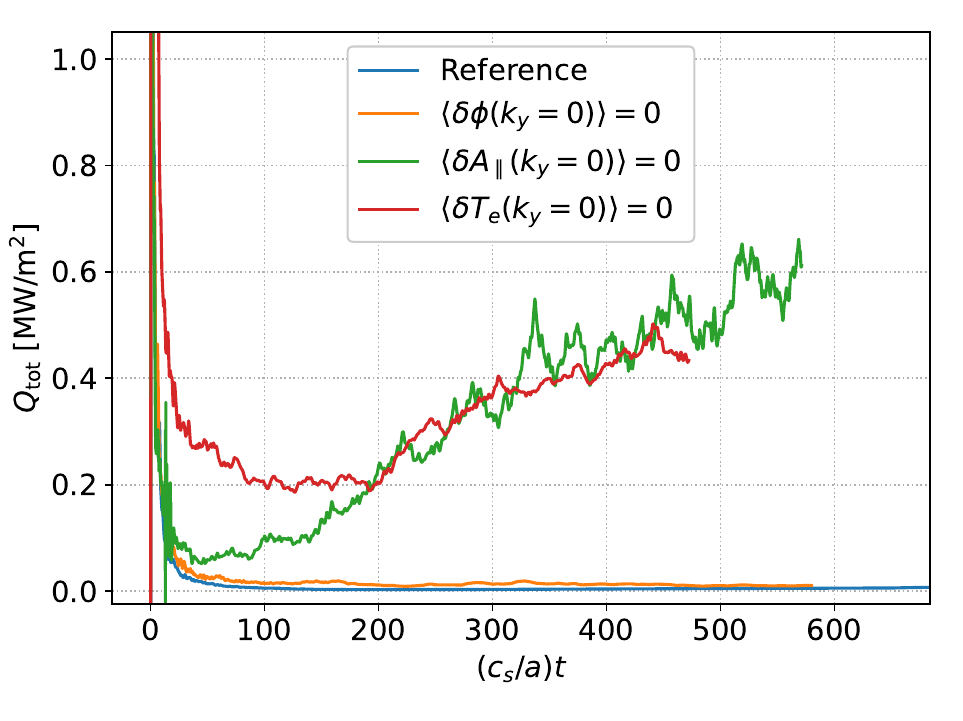}
    \caption{Time trace of the total heat flux from the nominal simulation (blue line) and three test simulations where the zonal flows (orange line), zonal fields (green line) or zonal temperature (red line) are removed from the system.}
    \label{fig:saturation}
\end{figure}

Previous theories and simulations of microtearing turbulence have reported various saturation processes through energy transfer to long and short wavelengths~\cite{doerk2011,drake1980}, background shear flow~\cite{guttenfelder2011}, zonal fields~\cite{pueschel2020,giacomin2023a}, electron temperature flattening~\cite{ajay2022}, and cross-scale interaction~\cite{maeyama2017}. In the STEP MTM simulations considered here, we find that electron temperature flattening and zonal fields play an important role in the saturation mechanism. As the electrons move swiftly along the perturbed magnetic field lines associated with magnetic islands at the rational surfaces, they undergo a periodic radial excursion, leading to a flattening of the electron temperature. Given that the electron temperature gradient provides the drive for the MTM instability, this flattening can locally stabilise the mode. We test whether this occurs in our simulations by removing the zonal electron temperature perturbations that cause the local temperature flattening, i.e. by redefining the zonal component of the electron distribution function as $\langle\delta f_e^\mathrm{mod}\rangle_{y, \theta} = \langle\delta f_e\rangle_{y, \theta} - K(r)[m_ev^2/(2T_e)-1.5]\langle F_{e0}\rangle_{y, \theta}$, where $\langle \cdot\rangle_{y, \theta}$ denotes the flux-surface average, $F_{e0}$ is the electron background Maxwellian distribution function and $K(r)$ is a function of the radial coordinate only, which is set at each time step such that $\langle\delta T_e\rangle_{y,\theta}=0$~\cite{ajay2022}. The resulting simulation, shown in Figure~\ref{fig:saturation}, returns a much larger heat flux than the reference case. 
Zonal fields can also provide a strong saturation mechanism for MTM turbulence by reducing the amplitude of non-zonal $\delta A_\parallel$ modes and the subsequent magnetic stochasticity. We also perform a nonlinear test simulation where the zonal fields are removed. As shown in Figure~\ref{fig:saturation}, this test simulation also returns much larger heat flux than the nominal simulation. Interestingly, the time trace of the heat flux from the two test simulations agrees very well, thus suggesting a competition (or a link) between zonal fields and local temperature flattening as saturation mechanisms. 
An additional simulation test also in Fig.~\ref{fig:saturation} shows that zonal flows are not relevant for MTM turbulence saturation, at least in the case considered here.

\section*{References}
\bibliographystyle{unsrt}
\bibliography{bibliography}

\end{document}